\documentclass[aps,superscriptaddress,amsmath,amssymb]{revtex4}
\usepackage{amsmath}
\usepackage{graphicx}
\usepackage{color}
\usepackage{dcolumn}
\usepackage{bm}
\usepackage{slashed}
\usepackage[colorlinks,linkcolor=red,anchorcolor=green,citecolor=blue,CJKbookmarks=True]{hyperref}

\newcommand{\bea}{\begin{eqnarray*}}
\newcommand{\eea}{\end{eqnarray*}}
\newcommand{\bean}{\begin{eqnarray}}
\newcommand{\eean}{\end{eqnarray}}
\newcommand{\eqs}[1]{Eqs.(\ref{#1})}
\newcommand{\eq}[1]{Eq.(\ref{#1})}
\newcommand{\meq}[1]{(\ref{#1})}
\newcommand{\grad}{\nabla}

\newcommand{\non}{\nonumber \\}

\newcommand{\hsp}{\hspace{0.1mm}}

\newcommand{\pp}{\partial}

\begin{document}
\title{Gauge Invariant Perturbations of General Spherically Symmetric Spacetimes}

\author{Wentao Liu}
\affiliation{Department of Physics, Key Laboratory of Low Dimensional Quantum Structures and Quantum Control of Ministry of Education, and Synergetic Innovation Center for Quantum Effects and Applications, Hunan Normal
University, Changsha, Hunan 410081, P. R. China}

\author{Xiongjun Fang}
\email[Corresponding author: ]{fangxj@hunnu.edu.cn} \affiliation{Department of Physics, Key Laboratory of Low Dimensional Quantum Structures and Quantum Control of Ministry of Education, and Synergetic Innovation Center for Quantum Effects and Applications, Hunan Normal
University, Changsha, Hunan 410081, P. R. China}
\affiliation{GCAP-CASPER, Department of Physics, Baylor University, Waco, Texas 76798-7316, USA}

\author{Jiliang Jing}
\affiliation{Department of Physics, Key Laboratory of Low Dimensional Quantum Structures and
Quantum Control of Ministry of Education, and Synergetic Innovation Center for Quantum Effects and Applications, Hunan Normal
University, Changsha, Hunan 410081, P. R. China}

\author{Anzhong Wang}
\affiliation{GCAP-CASPER, Department of Physics, Baylor University, Waco, Texas 76798-7316, USA}

\begin{abstract}

In this paper, the gauge choices in general spherically symmetric spacetimes are explored. In particular, we construct the gauge invariant variables and the master equations for both the Detweiler easy gauge and the Regge-Wheeler gauge, respectively. The particular cases for $l=0,1$ are also investigated. Our results provide analytical calculations of metric perturbations in general spherically symmetric spacetimes, which can be applied to various cases, including the effective-one-body problem. A simple example is presented to show how the metric perturbation components are related to the source perturbation terms.

~\\
\textbf{metric perturbation, gauge invariant, master equation}

~\\
\textbf{PACS number(s)}: 04.25.Nx, 04.30.Db, 04.70.-s

\end{abstract}

\maketitle

\section{Introduction}

Metric perturbations of spacetimes are an important issue. The solution of the Einstein field equations (EFEs) for static, vacuum and spherically symmetric spacetime is the Schwarzschild spacetime. And the metric perturbations of the Schwarzschild black hole have been studied for a long time. To begin with, Regge, Wheeler and Vishveshwara et.al studied the odd-parity perturbation \cite{ReggeWheeler1957, Vishveshwara1970PRD}, while Zerilli and Moncrief investigated the even-parity perturbation \cite{Zerilli1970PRL,Zerilli1970PRD,Moncrief1974ext}. The perturbation theory of the Schwarzschild spacetime has been well summarized in Chandrasekhar's monograph \cite{chandrasekharbook}. After decades of research and development, this theory can be applied to a variety of different physical problems. A useful application is the quasi-normal modes of the perturbed black holes, which was initiated by Vishveshwara \cite{Vishveshwara1970Nature}, Chandrasekhar \cite{chandrasekhar1975} and Mashhoon \cite{Mashhoon1984} et.al., and the review articles of this topic can be found in \cite{Kokkotas1999,Nollert1999,Cardoso2004, Cardoso2009CQG,Konoplya2011,Pani2013IJMPA}. Another application is studying a particle moving around the Schwarzschild black hole. One can treat this point-particle as a perturbation of the Schwarzschild spacetime \cite{Price1971,Martel2004}. In addition, studying the metric perturbation can promote the analysis of the stability of the Schwarzschild spacetime \cite{Price1972,Dafermos2016,Wald2018}.

In perturbation theory in general relativity, the redundant coordinate freedom can be eliminated by choosing specific gauges. The most familiar gauge in the Schwarzschild spacetime is the Regge-Wheeler (RW) gauge, which was first presented by Regge and Wheeler \cite{ReggeWheeler1957}. And where they also analysed the spherical harmonics and decomposed the general perturbation in the Schwarzschild spacetime into odd-parity and even-parity sectors. The RW gauge has the obvious advantage of algebraic simplicity, and it is widely used in the literature. Since then, the construction and the physical meaning of the gauge-invariant properties have attracted lots of attention. Using Lagrangian and Hamiltonian variational principles for the perturbation, Moncrief considered that the metric perturbations can be decomposed into the gauge invariant part and the gauge dependent part \cite{Moncrief1974ext,Moncrief1974int}. Gerlach and Sengupta discussed the construction of gauge invariant properties in general spherically symmetric spacetimes \cite{Gerlach1979,Gerlach1980}. Thorne reviewed and summarized various scalar, vector and tensor spherical harmonics with a uniform notation \cite{Thorne1980}. Martel and Poisson presented a gauge-invariant and covariant formalism, and also showed that the energy or angular-momentum radiation can be expressed in terms of gauge-invariant scalar functions \cite{Poisson2005}. Recently, Sopuerta considered that the master functions are linear combinations of the metric perturbations and their first-order derivatives, and discussed about the master equation for vacuum spherically symmetric spacetimes \cite{Sopuerta2021}. Besides the RW gauge, there exists a variety of gauge choices. For example, the light-cone gauge, which presented by Preston and Poisson \cite{Poisson2006}, can provide geometrical meaning to the coordinates in perturbed spacetimes. Another gauge choice is named as easy (EZ) gauge \cite{Thompson2016}, which was devised by Detweiler when he considered the gravitational self-force problem in the perturbed Schwarzschild spacetime \cite{Detweiler2008}. In the EZ gauge, the metric perturbation is singular on the black-hole horizon \cite{Poisson2018}.

Generally speaking, for metric perturbation of a spherically symmetric spacetime, the standard process is to decouple the even-parity and the odd-parity EFEs, obtain the wave equations and then solve the one dimensional Schr\"{o}dinger-like equation with an effective potential. One of the most important step is to construct the gauge-invariant variable and obtain the master equation. When the background metric takes the form
\begin{equation}
\label{ABeq1}
ds^2=-f(r)dt^2+f(r)^{-1}dr^2+ r^2(d\theta^2+\sin^2\theta d\varphi^2),
\end{equation}
the problems have been thoroughly studied for the theories of Einstein \cite{Kodama2003a,Kodama2003b}, Einstein-Maxwell  \cite{Kodama2004,Kodama2011}, and  Lovelock  \cite{TS09,TS10,Kodama2011} in higher dimensions. However, if one considers a non-vacuum spherical static black hole with hairs \cite{Heisenberg2018,Myung2019,Tomikawa2021}, the metric in general cannot be cast in the above form.  Instead,  the most general spherically symmetric static spacetimes should be described by the metric
\begin{equation}
\label{ABneq1}
ds^2=-A(r)dt^2+B(r)dr^2+r^2(d\theta^2+\sin^2\theta d\varphi^2),
\end{equation}
where $A\cdot B\neq 1$, for which perturbations have not been studied in detail so far. In particular, based on the Post-Newtonian (PN) approximation, Darmour et.al investigated the gravitational radiation generated by inspiralling compact binary systems and presented a novel approach to map the two-body problem onto an effective-one-body (EOB) system \cite{Damour1999,Damour2000}. Recently, the discussions of self-consistent of radiation-reaction force in the EOB system shows that one should first solve the gravitational perturbation in the most general shyerically symmetric spacetimes \cite{Jing2022}. Therefore, a natural question is how to construct  gauge-invariant perturbation variables in the most general spherically symmetric spacetimes (\ref{ABneq1}), and then study the even-parity and odd-parity perturbations.

With the above considerations as our main motivations, in this paper we consider the most general spherically symmetric background spacetimes with metric perturbations, and the construction of gauge invariant variables. For even-parity perturbations, we find that there exist several gauge choices, including the EZ gauge and the RW gauge. Under the EZ gauge, we construct the gauge-invariant variables and obtain a third-order master equations. However, the third-order equation can be written as a second-order equation after the separation of radial and time variables. Under the RW gauge, a similar situation also occurs. For the odd-parity perturbations, the master equation remains a second-order wave equation as usual. It should be noted that such developed formulas are not only applicable to the most general EOB system, as pointed above  \cite{Jing2022}, but also to other modified theories of gravity, in which the background is described by the
most general metric (\ref{ABneq1}). These include  theories with high-order derivative terms \cite{Bert18a,Bert18b,SJ22}. In such theories, the field equations can be always written as $G_{\mu\mu} = \kappa T^{\text{eff.}}_{\mu\nu}$, where $T^{\text{eff.}}_{\mu\nu}$ represent the modifications to general relativity (GR). Certainly, in such theories extra fields are often introduced. In the latter, we need to consider
not only the effective Einstein field equations, but also the equations for matter fields. In this paper we shall mainly focus on the effective Einstein field equations [cf. Eq.(\ref{eq16}) to be given below and other components given in Appendix B], that is, the perturbations of the most general
spherically symmetric metric, and leave the  studies of perturbations for matter field equations to another occasion, as the latter will be involved with specific modified theories.

Considering the general metric perturbations in static spherically symmetric spacetimes, Throne showed how to construct a ten-spherical-harmonic basis \cite{Thorne1980}. Through this paper, we use the A-K notation \cite{Thompson2016}, which dealt only with the Schwarzschild spacetime as the background, and was first presented by Detweiler when he considered the self-force problem. The advantage of using this notation is that one can find the relation between the metric perturbation components and the gauge invariants. In this paper, through the gauge invariants representing different combinations of the metric components, we show that the gauge-invariant variables have the similar structure under the EZ and RW gauges.

The rest of this paper is organized as follows. In Sect. II, we discuss the basic framework. First, the ten orthogonal harmonics basis are introduced. Then the decomposition of non-vacuum Einstein equations and the A-K notation are reviewed. After the investigation of gauge freedom, we consider the EZ and RW gauges. In Sect. III, we first consider the gauge invariant properties. Then we focus on constructing the master equations for both even-parity and odd-parity perturbations. We also study the cases for $l=0,1$. In Sect. IV, an example that a small particle goes around a circular orbit in spherically symmetric spacetimes is investigated. Finally, we summarize our main results with some discussions.

Throughout this paper, we use the A-K notation similarly to \cite{Thompson2016}. Units will be chosen in which $c=G=1$. In Sect. III, the subscript, e.g., $\psi_0$ and $\psi_1$, always represents quantities for $l=0$ and $l=1$ cases, respectively. And the superscript with Roman letters, i.e. $\chi^\mathrm{I}$ and $\chi^\mathrm{II}$, represent the quantities under the EZ gauge or under the RW gauge, respectively.

\section{Basic Framework}

\subsection{Orthogonal Harmonics Basis}

Let us start with the most general spherically symmetric spacetimes
\begin{equation}
\label{SO2spacetime}
ds^2=g_{ab}^{(0)}dx^adx^b=-e^{2\Phi(r)}dt^2+e^{2\Lambda(r)}dr^2+r^2(d\theta^2+\sin^2\theta d\varphi^2) .
\end{equation}
To decompose tensor fields on the above background, we choose the orthogonal basis composed of scalar spherical harmonics, vector harmonics and tensor harmonics. First, we define two unnormalized and orthogonal co-vectors $v$ and $n$
\begin{equation}
v_a=(-1,0,0,0), \quad\quad n_a=(0,1,0,0),
\end{equation}
the projection operator onto the sphere surface
\begin{equation}
\Omega_{ab}=g_{ab}^{(0)}+e^{2\Phi}v_a v_b-e^{2\Lambda}n_a n_b=r^2 \text{diag}(0,0,1,\sin^2\theta),
\end{equation}
and the spatial Levi-Civita tensor, $\epsilon_{abc}\equiv v^d\epsilon_{dabc}$, where $\epsilon_{tr\theta\phi}=e^{\Phi+\Lambda}r^2\sin\theta$.

In general, the complete basis on the 2-sphere are constructed by 1-scalar spherical harmonic, $Y^{lm}=Y^{lm}(\theta,\varphi)$, 3 pure-spin vector harmonics and 6 tensor harmonics \cite{Thorne1980}. The pure-spin vector harmonics are given by
\begin{equation}
Y_a^{E,lm}=r\grad_a Y^{lm},\quad Y_a^{B,lm}=r\epsilon_{ab}\hsp^c n^b\grad_c Y^{lm}, \quad Y_a^{R,lm}=n_aY^{lm}.
\end{equation}
And the pure-spin tensor harmonics are given by
\begin{align}
T_{ab}^{T0,lm}=\Omega_{ab}Y^{lm}, &\quad \quad T_{ab}^{L0,lm}=n_an_b Y^{lm}, \\
T_{ab}^{E1,lm}=rn_{(a}\grad_{b)}Y^{lm}, &\quad \quad T_{ab}^{B1,lm}=rn_{(a}\epsilon_{b)c}\hsp^d n^c\grad_d Y^{lm}, \\
T_{ab}^{E2,lm}=r^2\left(\Omega_a\hsp^c\Omega_b\hsp^d-\frac{1}{2}\Omega_{ab}\Omega^{cd}\right)\grad_c\grad_dY^{lm}, &\quad \quad
T_{ab}^{B2,lm}=r^2\Omega_{(a}\hsp^c\epsilon_{b)e}\hsp^d n^e\grad_c\grad_d Y^{lm}.
\end{align}
Note that the vector harmonics are orthogonal to each other
\begin{equation}
\oint Y_a^{A,lm}(Y^a_{A',l'm'})^*d\Omega =N^{(vec)}(A,r,l)\delta_{AA'}\delta_{ll'}\delta_{mm'} ,
\end{equation}
with $\{A,A'\}=\{E,B,R\}$ and $N^{(vec)}(A,r,l)$ is the specific normalization factor for vector harmonics. The tensor harmonics are also orthogonal to each other
\begin{equation}
\oint T_{ab}^{A,lm}(T^{ab}_{A',l'm'})^*d\Omega =N^{(ten)}(A,r,l)\delta_{AA'}\delta_{ll'}\delta_{mm'} ,
\end{equation}
with $\{A,A'\}=\{T0, L0,E1,E2,B1,B2\}$ and $N^{(ten)}(A,r,l)$ is the specific normalization factor for tensor harmonics. The expression for these normalization functions $N^{(vec)}$ and $N^{(ten)}$ are given in Appendix A.

\subsection{Decomposition of Linearized Einstein Equations}
For perturbed spacetimes, we use $h_{ab}$ to represent the linear perturbation of the background spacetime $g_{ab}^{(0)}$, i.e., the metric of the perturbed spacetime can be written as
\begin{equation}
g_{ab}=g_{ab}^{(0)}+h_{ab}.
\end{equation}
The background metric and the perturbed metric satisfied the Einstein Field Equations (EFEs)
\begin{align}
 G_{ab}(g^{(0)}) &= 8\pi T_{ab}, \\
 G_{ab}(g^{(0)}+h) &= 8\pi (T_{ab}+\mathcal{T}_{ab}),
\end{align}
where $T_{ab}$ and $\mathcal{T}_{ab}$ denote the non-vacuum background and the perturbed energy-momentum tensor, respectively. Expanding the EFEs in terms of $h_{ab}$, we get
\begin{equation}
 G_{ab}(g^{(0)}+h)=G_{ab}(g^{(0)})-\frac{1}{2}E_{ab}(h),
\end{equation}
where $E_{ab}$ is the linearized Einstein operator
\begin{align}
\label{PerEFEs}
E_{ab}(h) =& \Box h_{ab}+\nabla_a\nabla_bh^{c}_{~c}-2\nabla_{(a}\nabla^ch_{b)c}+2R^{~c~d}_{a~b}h_{cd}
-(R_a\hsp^c h_{bc}+R_b\hsp^c h_{ac}) \non
& +g_{ab}(\nabla^c\nabla^dh_{cd}-\Box h^{d}_{~d})-g_{ab}R^{cd}h_{cd}+R h_{ab} \non
=& -16\pi\mathcal{T}_{ab}.
\end{align}
Note that now the Ricci curvature $R_{ab}$ and the scalar curvature $R$ of the background do not vanish in general. If $R_{ab}$ and $R$ vanish, then the background metric would reduce to the Schwarzschild metric, which is the situation discussed in \cite{Thompson2016}.

Detweiler decomposed the harmonic modes of the perturbed metric $h_{ab}$ as
\begin{align}
\label{eq17}
h_{ab}^{lm} =& \mathrm{A~}v_av_bY^{lm}+2\mathrm{B~}v_{(a}Y_{b)}^{E,lm}+2\mathrm{C~}v_{(a}Y_{b)}^{B,lm}+2\mathrm{D~}v_{(a}Y_{b)}^{R,lm}+\mathrm{E~}T^{T0,lm}_{ab} \non
& +\mathrm{F~}T_{ab}^{E2,lm}+\mathrm{G~}T^{B2,lm}_{ab}+2\mathrm{H~}T^{E1,lm}_{ab}+2\mathrm{J~}T^{B1,lm}_{ab}+\mathrm{K~}T^{L0,lm}_{ab} ,
\end{align}
where all coefficients A through K are scalar functions of $(t,r)$, which were referred to as the A-K coefficients in \cite{Thompson2016}.

For even-parity (polar part) perturbations with $l\geq 2$, the perturbed metric can be decomposed as
\begin{equation}
h_{ab}^{\text {even }}=\left(\begin{array}{cccc}
	\mathrm{A~}Y_{lm} & -\mathrm{D~}Y_{lm} & -r\mathrm{B~}\partial_{\theta} Y_{lm} & -r\mathrm{B~}\partial_{\phi} Y_{lm} \\
	\text {Sym} & \mathrm{K~}Y_{lm} & r\mathrm{H~}\partial_{\theta} Y_{lm} & r\mathrm{H~}\partial_{\phi} Y_{lm} \\
	\text {Sym} & \text {Sym} & r^{2}\left[\mathrm{E}+\mathrm{F}\left(\partial_{\theta}^2+\frac{1}{2}l(l+1)\right)\right]Y_{lm} & r^{2}\mathrm{F}\left[\partial_{\theta}\partial_{\phi}-\cot\theta\partial_{\phi}\right]Y_{lm} \\
	\text {Sym} & \text {Sym} & \text {Sym} & r^{2} \sin^2\theta\left[\mathrm{E}-\mathrm{F}\left(\partial_{\theta}^{2}+\frac{1}{2} l(l+1)\right)\right]Y_{lm}
\end{array}\right) .
\end{equation}
And for the odd-parity (axial part) perturbations with $l\geq 2$, the perturbed metric can be decomposed as
\begin{equation}
h_{a b}^{\mathrm{odd}}=e^{-\Phi-\Lambda}
\left(\begin{array}{cccc}
	0 & 0 &  r \csc\theta \mathrm{C~}\partial_{\phi} Y_{lm} & -r \sin \theta\mathrm{~C~}\partial_{\theta} Y_{lm} \\
	0 & 0 & -r \csc\theta \mathrm{J~}\partial_{\phi} Y_{lm} & r \sin \theta \mathrm{~J~}\partial_{\theta} Y_{lm} \\
	\mathrm{Sym} & \mathrm{Sym} & -r^2\csc\theta \mathrm{G}\left[\partial_{\theta} \partial_{\phi}-\cot \theta \partial_{\phi}\right] Y_{lm} & -\frac{r^{2}}{2}\mathrm{G}\left[\csc\theta \partial_{\phi}^{2}+\cos\theta\partial_{\theta}-\sin\theta\partial_{\theta}^2\right]Y_{lm} \\
	\mathrm{Sym} & \mathrm{Sym} & \mathrm{Sym} & r^{2}\mathrm{G}\left[\sin \theta \partial_{\theta} \partial_{\phi}-\cos \theta \partial_{\phi}\right] Y_{lm}
\end{array}\right) .
\end{equation}
These A-K notations could be linearly transformed into the notation taken by Regge and Wheeler \cite{Thompson2016}. To get the explicit expression of coefficients A-K, one can project the metric perturbation onto the tensor harmonic basis. We present these coefficients in Appendix A.

Now, we can get the A-K components of any rank-2 tensor in the spherically symmetric background. For example, from \eq{PerEFEs}, one can write
\begin{equation}
\label{eq16}
-16\pi\mathcal{T}_{\mathrm{A}}=E_{\mathrm{A}}=e^{4\Phi}\oint E_{ab}(h)(v^av^bY^{\ast}_{lm})d\Omega ,
\end{equation}
to represent the A-term of $\mathcal{T}_{ab}$ or $E_{ab}(h)$, which along the direction of $v^av^bY^{\ast}_{lm}$. The expression for $E_{\mathrm{A}}$-$E_{\mathrm{K}}$ are given in Appendix B.

\subsection{The Gauge Transformation}
Under a gauge transformation, $\tilde x^a=x^a+\xi^a$, the first-order metric perturbation $h_{ab}$ would be transformed as
\begin{equation}
\tilde h_{ab}=h_{ab}-2\grad_{(a}\xi_{b)} .
\end{equation}
As a vector, $\xi^a$ can be projected onto the pure-spin harmonic basis as
\begin{equation}
\xi_a=\mathrm{P~}v_aY_{lm}+\mathrm{R~}Y_a^{R,lm}+\mathrm{S~}Y_a^{E,lm}+\mathrm{Q~}Y_a^{B,lm} ,
\end{equation}
where $\mathrm{P}$, $\mathrm{R}$, $\mathrm{S}$ and $\mathrm{Q}$ are scalar functions of $(t,r)$. The functions $\mathrm{P}$, $\mathrm{R}$ and $\mathrm{S}$ describe three degrees of gauge freedom for even-parity perturbations, while the function $\mathrm{Q}$ describes one degree of the gauge freedom for odd-parity perturbations. Next we use $\Delta$ to represent the A-K projections of $2\grad_{(a}\xi_{b)}$. For example,
\begin{equation}
\Delta \mathrm{A} = \mathrm{A}-\mathrm{\tilde A} = 2e^{4\Phi}\oint \grad_{(a}\xi_{b)}(v^av^bY^{\ast}_{lm})d\Omega ,
\end{equation}
here A and $\mathrm{\tilde A}$ correspond to the projections of $h_{ab}$ and $\tilde h_{ab}$, respectively. Projecting $2\grad_{(a}\xi_{b)}$ onto the tensor harmonic basis, we obtain the A-K components of the term $2\grad_{(a}\xi_{b)}$,
\begin{align}
\label{DeltaAK}
& \Delta \mathrm{A} = -2\frac{\pp \mathrm{P}}{\pp t}-2e^{-2\Lambda+2\Phi}\frac{\pp \Phi}{\pp r}\mathrm{R},
& \Delta \mathrm{B} &= -\frac{\pp \mathrm{S}}{\pp t}+\frac{1}{r}\mathrm{P}, \non
& \Delta \mathrm{C} = -\frac{\pp }{\pp t}\mathrm{Q},
& \Delta \mathrm{D} &= \left(\frac{\pp }{\pp r}-2\frac{\pp \Phi}{\pp r} \right)\mathrm{P} -\frac{\pp \mathrm{R}}{\pp t},  \non
& \Delta \mathrm{E} = 2\frac{e^{-2\Lambda}}{r}\mathrm{R}-\frac{l(l+1)}{r}\mathrm{S},
& \Delta \mathrm{F} &= \frac{2}{r}\mathrm{S}, \non
& \Delta \mathrm{G} = \frac{2}{r}\mathrm{Q}, &\Delta \mathrm{H} &= \frac{1}{r}\mathrm{R}+\left(\frac{\pp}{\pp r}-\frac{1}{r}\right)\mathrm{S}, \non
& \Delta \mathrm{J} = \left(\frac{\pp}{\pp r}-\frac{1}{r}-\frac{\pp \Phi}{\pp r}-\frac{\pp \Lambda}{\pp r}\right)\mathrm{Q},
& \Delta \mathrm{K} &=\left(2\frac{\pp }{\pp r}-2\frac{\pp \Lambda}{\pp r}\right)\mathrm{R}.
\end{align}

\subsection{Gauge Choices}
Generally speaking, $\xi^a$ has four independent functions, representing four degrees of freedom in spherically symmetric spacetimes. Hence by properly choosing these four functions we can work with different gauges. For example, under the gauge transformation, the scalar function $\mathrm{F}(t,r)$ would transform as
\begin{equation}
\mathrm{\tilde F} = \mathrm{F}-\Delta \mathrm{F} = \mathrm{F}-\frac{2}{r}\mathrm{S}.
\end{equation}
By setting $\mathrm{S}=r\mathrm{F}/2$, one degree of the gauge freedom is fixed, and $\mathrm{\tilde F}=0$. Then substituting this S back into \eq{DeltaAK}, one can move on to eliminate the next degrees of freedom. In even-parity perturbations, properly choosing the functions $\mathrm{P}$, $\mathrm{R}$ and $\mathrm{S}$ would fix three variables of the metric perturbation. In the odd-parity properly choosing the function $\mathrm{Q}$, one can fix one variable of the metric perturbation.

Note that in the odd-parity sector, $\Delta \mathrm{G}$ is proportional to $\mathrm{Q}$, but there exists some derivative relation between $\Delta \mathrm{C}$, $\Delta \mathrm{J}$ and Q. If we want to eliminate $\mathrm{\tilde G}$ under the gauge transformation, just set $Q=r\mathrm{G}/2$ then $\mathrm{\tilde G}=0$ and $\Delta \mathrm{C}$ and $\Delta \mathrm{J}$ are uniquely determined. If we eliminate $\mathrm{\tilde C}$ under the gauge transformation rather than $\mathrm{\tilde G}$, then $\mathrm{Q}$ could be an arbitrary function of $r$ with some integration constant, which could not determine $\Delta \mathrm{G}$ and $\Delta \mathrm{J}$ completely. Such a choice cannot completely fix the gauge freedom. Similarly, in the even-parity sector, in order to fix the gauge completely, one may first fix $\mathrm{S}$ from $\Delta \mathrm{F}$. Afterward, there are still several choices to fix $\mathrm{P}$ and $\mathrm{R}$ via $\Delta \mathrm{A}$, $\Delta \mathrm{B}$, $\Delta \mathrm{E}$ or $\Delta \mathrm{H}$. Below, we would discuss two useful gauge choices.

\textbf{Regge-Wheeler Gauge.} Regge and Wheeler first presented the RW gauge by setting certain RW variables to zero \cite{ReggeWheeler1957}. In the even-parity sector of the Schwarzschild spacetime, to eliminate three gauge freedom, RW set
\begin{equation}
h_0^{RW,\text{even}}=h_1^{RW,\text{even}}=G^{RW}=0 ,
\end{equation}
which corresponds to setting $\mathrm{\tilde B}=\mathrm{\tilde F}=\mathrm{\tilde H}=0$ in the A-K notation. And in the odd-parity sector, RW set $h_2^{RW}=0$, which corresponds to setting $\mathrm{\tilde G}=0$ \cite{Thompson2016}. For general spherically symmetric spacetimes, we set
\begin{align}
& \mathrm{S}^{RW} = \frac{r}{2}\mathrm{F}, & \mathrm{P}^{RW} &= r \mathrm{B}+\frac{r^2}{2}\frac{\pp \mathrm{F}}{\pp t}, \non
& \mathrm{R}^{RW} = r \mathrm{H}-\frac{r}{2}\frac{\pp \mathrm{F}}{\pp r}, & \mathrm{Q}^{RW} &= \frac{r}{2}\mathrm{G},
\end{align}
which means
\begin{equation}
\xi_a^{RW}=\left(r \mathrm{B}+\frac{r^2}{2}\frac{\pp \mathrm{F}}{\pp t}\right)v_aY_{lm}
+\left(r \mathrm{H}-\frac{r}{2}\frac{\pp \mathrm{F}}{\pp r}\right)Y_a^{R,lm}+\frac{r}{2}\mathrm{F}Y_a^{E,lm}+\frac{r}{2}\mathrm{G}Y_a^{B,lm} .
\end{equation}
It should note that this gauge choice is exactly the same as the gauge choice with the A-K notation in the Schwarzschild spacetime, see, for example, Eq.(6.7) in \cite{Thompson2016}.

\textbf{Easy gauge.} The EZ gauge was first introduced by Detweiler when he considered the self-force problem, in which the following metric components are set to zero,
\begin{equation}
\mathrm{\tilde B}=\mathrm{\tilde E}=\mathrm{\tilde F}=\mathrm{\tilde G}=0 .
\end{equation}
To eliminate these metric components, the components of the gauge vector are chosen as
\begin{align}
& \mathrm{S}^{EZ} = \frac{r}{2}\mathrm{F}, & \mathrm{P}^{EZ} &= r \mathrm{B}+\frac{r^2}{2}\frac{\pp \mathrm{F}}{\pp t}, \non
& \mathrm{R}^{EZ} = \frac{r}{4}l(l+1)e^{2\Lambda} \mathrm{F}+\frac{r}{2}e^{2\Lambda} \mathrm{E}, & \mathrm{Q}^{EZ} &= \frac{r}{2}\mathrm{G},
\end{align}
which means
\begin{equation}
\xi_a^{EZ}=\left(r \mathrm{B}+\frac{r^2}{2}\frac{\pp \mathrm{F}}{\pp t}\right)v_aY_{lm}
+\left(rl(l+1)e^{2\Lambda} \mathrm{F}+\frac{r}{2}e^{2\Lambda} \mathrm{E}\right)Y_a^{R,lm}+\frac{r}{2}\mathrm{F}Y_a^{E,lm}+\frac{r}{2}\mathrm{G}Y_a^{B,lm} .
\end{equation}

\section{Gauge Invariants and Master Equations}
In this section, we would first introduce a general set of gauge invariants in the spherically symmetric backgrounds. With these gauge invariants, we shall investigate how to construct the single master equation under certain gauge choices.

\subsection{Gauge Invariants}
Generally speaking, under any arbitrary gauge transformation, the gauge invariants can be constructed from \eq{DeltaAK}. For example,
\begin{align}
\Delta \mathrm{G} &= \frac{2}{r}\mathrm{Q}, \\
\Delta \mathrm{J} &= \left(\frac{\pp}{\pp r}-\frac{1}{r}-\frac{\pp \Phi}{\pp r}-\frac{\pp \Lambda}{\pp r}\right)\mathrm{Q} ,
\end{align}
from which, we obtain
\begin{equation}
\Delta \mathrm{J}+\frac{r}{2}\left(\frac{\pp \Phi}{\pp r}+\frac{\pp\Lambda}{\pp r}-\frac{\pp }{\pp r} \right)\Delta \mathrm{G} = 0 .
\end{equation}
The above equation indicates that one can define
\begin{equation}
\alpha = \mathrm{J}+\frac{r}{2}\left(\frac{\pp \Phi}{\pp r}+\frac{\pp\Lambda}{\pp r}-\frac{\pp }{\pp r} \right)\mathrm{G},
\end{equation}
and $\alpha$ is a gauge invariant quantity. Note that there are seven even-parity metric components and three odd-parity metric components in the metric perturbation, while there are three even-parity components and one odd-parity component in $\xi^a$, which tell us that we can construct four independent even-parity gauge invariants and two independent odd-parity gauge invariants for $l\geq 2$ cases. Following the construction of $\alpha$, we find that they can be constructed as
\begin{align}
\label{gaugeinv}
\alpha &= \mathrm{J}+\frac{r}{2}\left(\frac{\pp \Phi}{\pp r}+\frac{\pp\Lambda}{\pp r}-\frac{\pp }{\pp r} \right)\mathrm{G} , \non
\beta &= -\mathrm{C}-\frac{r}{2}\frac{\pp }{\pp t}\mathrm{G} , \non
\chi &= \mathrm{H}-\frac{1}{2}e^{2\Lambda}\mathrm{E}-\frac{l(l+1)}{4}e^{2\Lambda}\mathrm{F}-\frac{r}{2}\frac{\pp }{\pp r}\mathrm{F} , \non
\psi &= \frac{1}{2}\mathrm{K} -\frac{r}{2}e^{2\Lambda}\frac{\pp \Lambda}{\pp r}\mathrm{E}-\frac{1}{2}e^{2\Lambda}\mathrm{E}
-\frac{r}{2}e^{2\Lambda}\frac{\pp }{\pp r}\mathrm{E}-\frac{r}{4}l(l+1)e^{2\Lambda}\frac{\pp \Lambda}{\pp r}\mathrm{F}-\frac{1}{4}l(l+1)e^{2\Lambda}\mathrm{F} -\frac{r}{4}l(l+1)e^{2\Lambda}\frac{\pp }{\pp r}\mathrm{F} , \non
\delta &= \mathrm{D}+\frac{r}{2}e^{2\Lambda}\frac{\pp }{\pp t}\mathrm{E}+\left(2r\frac{\pp \Phi}{\pp r} -1\right) B-r\frac{\pp }{\pp r}\mathrm{B}
-\frac{r^2}{2}\frac{\pp^2}{\pp t\pp r}\mathrm{F}-\left[ r-r^2\frac{\pp \Phi}{\pp r}-\frac{r}{4}l(l+1)e^{2\Lambda}\right]\frac{\pp}{\pp t}\mathrm{F} , \non
\epsilon &= -\frac{1}{2}\mathrm{A}-\frac{r}{2}e^{2\Phi}\frac{\pp \Phi}{\pp r}\mathrm{E}-r\frac{\pp }{\pp t}\mathrm{B}
-\frac{r}{4}l(l+1)e^{2\Phi}\frac{\pp \Phi}{\pp r}\mathrm{F}-\frac{r^2}{2}\frac{\pp^2}{\pp t^2}\mathrm{F} .
\end{align}
These relations are the same as Eq. (7.5) of \cite{Thompson2016} when the background is vacuum, which degenerates to the Schwarzschild spacetime.

From now on, we shall work in the coordinates $\tilde x^a$. And for the save of simplicity, all the tildes will be dropped from now on. Note that we use the superscript $\mathrm{I}$ or $\mathrm{II}$ to denote the quantities or parameters under the EZ gauge or the RW gauge, respectively. Adopting the specific EZ gauge, we have
\begin{equation}
\mathrm{B}=\mathrm{E}=\mathrm{F}=\mathrm{G}=0 .
\end{equation}
Then, the gauge invariants become
\begin{align}
\label{gaugeinvtometricEZ}
\alpha &= \mathrm{J}, & \beta &=-\mathrm{C}, \non
\chi^\mathrm{I} &= \mathrm{H}, & \psi^\mathrm{I} &= \frac{1}{2}\mathrm{K}, \non
\delta^\mathrm{I} &= \mathrm{D}, & \epsilon^\mathrm{I} &= -\frac{1}{2}\mathrm{A} .
\end{align}
Here $\alpha$ and $\beta$ are not superscripted because they are the same under the EZ and RW gauges. Similarly, adopting the specific RW gauge, we have
\begin{equation}
\mathrm{B}=\mathrm{F}=\mathrm{H}=\mathrm{G}=0,
\end{equation}
and the even-parity gauge invariants become
\begin{align}
\label{gaugeinvtometricRW}
\chi^\mathrm{II} &= -\frac{1}{2}e^{2\Lambda}\mathrm{E} ,
& \psi^\mathrm{II} &= \frac{1}{2}\mathrm{K}-\frac{1}{2}\left(r\Lambda'+1\right)e^{2\Lambda}\mathrm{E}
-\frac{r}{2}e^{2\Lambda}\frac{\pp}{\pp r}\mathrm{E} , \non
\delta^\mathrm{II} &= \mathrm{D}+\frac{r}{2}e^{2\Lambda}\frac{\pp}{\pp t}\mathrm{E} , &
\epsilon^\mathrm{II} &= -\frac{1}{2}\mathrm{A}-\frac{1}{2}r\Phi'e^{2\Phi}\mathrm{E} .
\end{align}

Using these gauges, each A-K projection of the linearized Einstein equations, i.e. \eqs{projEA}-\meq{projEJ}, can be rewritten as a combination of the gauge invariants listed in \eq{gaugeinv}. The results can be found in Appendix C. It is obvious that under the EZ gauge, the relationship between the gauge invariants and the perturbed metric components seems simpler, hence we first study the master equation under the EZ gauge.

\subsection{Master Equations for \texorpdfstring{$l\geq 2$}.}

Assuming that the perturbation of the stress-energy $\mathcal{T}_{ab}$ is known, i.e. $E_{\mathrm{A}}$-$E_\mathrm{K}$ are known quantities, now we look for the master equations in terms of gauge invariants.

\subsubsection{\textbf{Even-parity Perturbations and the EZ gauge}}

The Bianchi identities indicate that not all seven even-parity projection equations in Appendix C are independent. It has been shown that there are four independent gauge invariants. Noting the specific structure of the expressions of $E_{\mathrm{A}}^\mathrm{I}$, $E_{\mathrm{D}}^\mathrm{I}$, $E_{\mathrm{F}}^\mathrm{I}$, $E_{\mathrm{H}}^\mathrm{I}$ and $E_{\mathrm{K}}^\mathrm{I}$, we find that one can obtain
$\pp\delta^\mathrm{I}/\pp t$ from $\pp E_{\mathrm{D}}^\mathrm{I}/\pp t$, and $\epsilon^\mathrm{I}$ from $E_{\mathrm{A}}^\mathrm{I}$ or $E_{\mathrm{F}}^\mathrm{I}$. Substituting these relations into $2E_{\mathrm{H}}^\mathrm{I}+E_{\mathrm{K}}^\mathrm{I}$, after a large but tedious calculation, we find that the coupled partial differential equations for $\chi^\mathrm{I}$ and $\psi^\mathrm{I}$ can be constructed as

\begin{align}
\label{prchiEZ}
\frac{\pp}{\pp r}\chi^\mathrm{I} =& \frac{2e^{2\Lambda-2\Phi}r}{\tau^\mathrm{I} \eta^{\mathrm{I}}}
\left(\sigma^\mathrm{I}\frac{\pp^2}{\pp t^2}\chi^\mathrm{I}-2\frac{\pp^2}{\pp t^2}\psi^\mathrm{I}
+\frac{1}{2}e^{2\Lambda}r\frac{\pp }{\pp t}E_\mathrm{D}^\mathrm{I} \right)
+\frac{\gamma^\mathrm{I}}{r\tau^\mathrm{I}}\chi^\mathrm{I}+\frac{\rho^\mathrm{I}}{r\tau^\mathrm{I}}\psi^\mathrm{I}\non
&-\frac{e^{2\Lambda}r}{2\tau^{\mathrm{I}}}\left(e^{2\Lambda}\lambda+2r\Phi'-2\right)E_\mathrm{F}^\mathrm{I}+\frac{e^{2\Lambda}r}{2\tau^\mathrm{I}}\left(E_\mathrm{K}^\mathrm{I}+2E_\mathrm{H}^\mathrm{I}\right) ,
\end{align}
\begin{align}
\label{prpsiEZ}
\frac{\pp}{\pp r}\psi^\mathrm{I} =& \frac{ e^{2\Lambda-2\Phi}r}{\tau^\mathrm{I}}
\left(\sigma^\mathrm{I}\frac{\pp^2}{\pp t^2}\chi^\mathrm{I}-2\frac{\pp^2}{\pp t^2}\psi^\mathrm{I}
+\frac{1}{2}e^{2\Lambda}r\frac{\pp}{\pp t}E_\mathrm{D}^\mathrm{I} \right)
+\frac{\mu^\mathrm{I}}{r \tau^\mathrm{I}}\chi^\mathrm{I}+\frac{\nu^\mathrm{I}}{r\tau^\mathrm{I}}\psi^\mathrm{I} \non
& +\frac{ e^{2\Lambda}r}{4\tau^\mathrm{I}}\left[\kappa^{\mathrm{I}}E^\mathrm{I}_{\mathrm{F}}+\eta^{\mathrm{I}}\left(E_\mathrm{K}^\mathrm{I}+2E_\mathrm{H}^\mathrm{I}\right)\right]
-\frac{1}{4}e^{4\Lambda-2\Phi}rE_\mathrm{A}^\mathrm{I} ,
\end{align}
where $\lambda=l(l+1)$, and $\sigma^\mathrm{I}$, $\tau^\mathrm{I}$, $\eta^\mathrm{I}$, $\gamma^\mathrm{I}$, $\rho^\mathrm{I}$, $\mu^\mathrm{I}$, $\nu^\mathrm{I}$, $\kappa^\mathrm{I}$ are functions determined by the background spacetime, which can be find in Appendix D. Note that in these two equations, there are no spatial derivatives of the source terms. Now the goal becomes to decouple the gauge invariants $\chi^\mathrm{I}$ and $\psi^\mathrm{I}$, e.g. \eqs{prchiEZ} and \meq{prpsiEZ}. Introducing the gauge invariant
\begin{equation}
\label{evenmastervEZ}
Z^\mathrm{I}\hsp^{(+)} = \sigma^\mathrm{I}\chi^\mathrm{I}-2\psi^\mathrm{I} ,
\end{equation}
we find that $Z^\mathrm{I}\hsp^{(+)}$ satisfies the master equation
\begin{equation}
\label{evenmasterEZ}
\left[\frac{2e^{2\Lambda-2\Phi}r}{N^{\mathrm{I}}\eta^{\mathrm{I}}}
\left(2r(\eta^\mathrm{I}-\sigma^\mathrm{I})\sigma^{\mathrm{I}}\frac{\pp^3}{\pp t^2 \pp r}
+\frac{N_\mathrm{T}^\mathrm{I}}{N^\mathrm{I}\eta^{\mathrm{I}}\tau^\mathrm{I}}\frac{\pp^2}{\pp t^2}\right)
+\frac{2r\sigma^{\mathrm{I}}\tau^\mathrm{I}}{N^{\mathrm{I}}}\frac{\pp^2}{\pp r^2}+\frac{N_\mathrm{R}^\mathrm{I}}{\left({N^\mathrm{I}}\right)^2}\frac{\pp}{\pp r}
+\frac{N_\mathrm{Z}^\mathrm{I}}{r\left({N^\mathrm{I}}\right)^2\tau^\mathrm{I}}\right]Z^\mathrm{I}\hsp^{(+)}
=S_{\text{even}}^\mathrm{I},
\end{equation}
with the source term given by
\begin{align}
\label{evensourceEZ}
S_{\text{even}}^\mathrm{I} =& \frac{e^{4\Lambda-2\Phi}r}{N^{\mathrm{I}}} \left[\frac{N_\mathrm{A}^\mathrm{I}}{2N^\mathrm{I}}E_\mathrm{A}^\mathrm{I}
+r\sigma^{\mathrm{I}}\tau^\mathrm{I}\frac{\pp}{\pp r}E_\mathrm{A}^\mathrm{I}
+\frac{rN_\mathrm{D}^\mathrm{I}}{N^\mathrm{I}\left(\eta^{\mathrm{I}}\right)^2\tau^\mathrm{I}}\frac{\pp}{\pp t}E_\mathrm{D}^\mathrm{I}
-\frac{2r^2(\eta^\mathrm{I}-\sigma^\mathrm{I})\sigma^{\mathrm{I}}}{\eta^{\mathrm{I}}}\frac{\pp^2}{\pp t\pp r}E_\mathrm{D}^\mathrm{I} \right]
-\frac{2e^{2\Lambda}r^2M^{\mathrm{I}}_{\mathrm{F}}\sigma^{\mathrm{I}}}{N^{\mathrm{I}}}\frac{\pp}{\pp r}E^{\mathrm{I}}_{\mathrm{F}} \non
& +\frac{e^{2\Lambda}r}{\left(N^{\mathrm{I}}\right)^2\tau^{\mathrm{I}}} \left[\frac{1}{2}N_\mathrm{F}^\mathrm{I}E_\mathrm{F}^\mathrm{I}
+N_\mathrm{HK}^\mathrm{I}\left(E_\mathrm{H}^\mathrm{I}+\frac{1}{2}E_\mathrm{K}^\mathrm{I}\right)\right]
-\frac{e^{2\Lambda}r^2\left(\eta^{\mathrm{I}}-\sigma^{\mathrm{I}}\right)\sigma^{\mathrm{I}}}{N^{\mathrm{I}}}
\left(2\frac{\pp}{\pp r}E^{\mathrm{I}}_{\mathrm{H}}+\frac{\pp}{\pp r}E^{\mathrm{I}}_{\mathrm{K}}\right) .
\end{align}
In the above equation, $N^\mathrm{I}$, $N_\mathrm{T}^\mathrm{I}$, $N_\mathrm{R}^\mathrm{I}$, $N_\mathrm{Z}^\mathrm{I}$ and $N_\mathrm{A}^\mathrm{I}$, $N_\mathrm{D}^\mathrm{I}$, $M^{\mathrm{I}}_\mathrm{F}$, $N_\mathrm{F}^\mathrm{I}$, $N_\mathrm{HK}^\mathrm{I}$ are functions depending only on the background. The explicit expressions of them can be found in Appendix D. Unlike the well-known Zerilli equation, the master equation under the EZ gauge is a third-order equation. However, rewriting \eq{evenmasterEZ} in the form
\begin{equation}
\left( a\frac{\pp^3}{\pp t^2\pp r}+b\frac{\pp^2}{\pp t^2}+c\frac{\pp^2}{\pp r^2}+d\frac{\pp}{\pp r}+e \right)Z^\mathrm{I}\hsp^{(+)}=S_{\text{even}}^\mathrm{I},
\end{equation}
setting $Z^\mathrm{I}\hsp^{(+)}=Z^\mathrm{I}\hsp^{(+)}(r)e^{i\omega t}$ and $S_{\text{even}}^\mathrm{I}=S_{\text{even}}^\mathrm{I}(r)e^{i\omega t}$, we obtain a second-order differential equation
\begin{equation}
\left[c\frac{\pp^2}{\pp r^2}+(d-a\omega^2)\frac{\pp}{\pp r}+(e-b\omega^2) \right]Z^\mathrm{I}\hsp^{(+)}(r)=S_{\text{even}}^\mathrm{I}(r) .
\end{equation}

If the background becomes the Schwarzschild spacetime, the master equation \meq{evenmasterEZ} will reduce to the result given in \cite{Thompson2016}, in other words, the result of Zerilli \cite{Zerilli1970PRL}. Further discussions about the degeneration of our result could be found in Appendix E.

Once \eq{evenmasterEZ} is solved, together with the definition of the master variable given by \eq{evenmastervEZ}, the gauge invariants $\chi^\mathrm{I}$ and $\psi^\mathrm{I}$ can be solved from
\begin{align}
\chi^\mathrm{I} =& \frac{1}{N^\mathrm{I}}
\left[\frac{4e^{2\Lambda-2\Phi}r^2\left(\eta^\mathrm{I}-\sigma^\mathrm{I}\right)}
{\eta^\mathrm{I}}\frac{\pp^2}{\pp t^2}Z^\mathrm{I}\hsp^{(+)}
+2r\tau^\mathrm{I}\frac{\pp }{\pp r}Z^\mathrm{I}\hsp^{(+)} +M_1^\mathrm{I}Z^\mathrm{I}\hsp^{(+)} -e^{4\Lambda-2\Phi}r^2\tau^\mathrm{I} E_\mathrm{A}^\mathrm{I} \right. \non
& \left. +\frac{2e^{4\Lambda-2\Phi}r^3\left(\eta^\mathrm{I}-\sigma^\mathrm{I}\right)}
{\eta^\mathrm{I}}\frac{\pp}{\pp t}E_\mathrm{D}^\mathrm{I}
+2e^{2\Lambda}r^2M^\mathrm{I}_{\mathrm{F}} E_\mathrm{F}^\mathrm{I}
+e^{2\Lambda}r^2\left(\eta^\mathrm{I}-\sigma^\mathrm{I}\right)\left(E_\mathrm{K}^\mathrm{I}+2E_\mathrm{H}^\mathrm{I}\right) \right] ,
\end{align}
\begin{align}
\psi^\mathrm{I} =& \frac{\sigma^\mathrm{I}}{2N^\mathrm{I}}
\left[\frac{4e^{2\Lambda-2\Phi}r^2\left(\eta^\mathrm{I}-\sigma^\mathrm{I}\right)}
{\eta^\mathrm{I}}\frac{\pp^2}{\pp t^2}Z^\mathrm{I}\hsp^{(+)}
+2r\tau^\mathrm{I}\frac{\pp }{\pp r}Z^\mathrm{I}\hsp^{(+)} -M_2^\mathrm{I}Z^\mathrm{I}\hsp^{(+)} -e^{4\Lambda-2\Phi}r^2\tau^\mathrm{I} E_\mathrm{A}^\mathrm{I} \right. \non
& \left. +\frac{2e^{4\Lambda-2\Phi}r^3\left(\eta^\mathrm{I}-\sigma^\mathrm{I}\right)}
{\eta^\mathrm{I}}\frac{\pp}{\pp t}E_\mathrm{D}^\mathrm{I}
+2e^{2\Lambda}r^2M^\mathrm{I}_{\mathrm{F}} E_\mathrm{F}^\mathrm{I}
+e^{2\Lambda}r^2\left(\eta^\mathrm{I}-\sigma^\mathrm{I}\right)\left(E_\mathrm{K}^\mathrm{I}+2E_\mathrm{H}^\mathrm{I}\right) \right],
\end{align}
where $M_1^\mathrm{I}$ and $M_2^\mathrm{I}$ are given in Appendix D. Then, the remaining two even-parity gauge invariants $\delta^\mathrm{I}$ and $\epsilon^\mathrm{I}$ can be obtained from \eqs{projEDgaugeinEZ} and \meq{projEFgaugeinEZ}, given respectively by
\begin{equation}
\delta^\mathrm{I} = \frac{1}{-2-e^{2\Lambda}\left(-2+\lambda\right)+4r\Lambda'}
\left(e^{2\Lambda}r^2E_\mathrm{D}^\mathrm{I}+e^{2\Lambda}\lambda r\frac{\pp}{\pp t}\chi^\mathrm{I}-4r\frac{\pp }{\pp t}\psi^\mathrm{I} \right) ,
\end{equation}
\begin{equation}
\epsilon^\mathrm{I} =e^{-2\Lambda+2\Phi}\left[\left(1-r\Lambda'+r\Phi'\right)\chi^\mathrm{I}
-\psi^\mathrm{I}+r\frac{\pp }{\pp r}\chi^\mathrm{I} +\frac{1}{2}e^{2\Lambda}r^2E_\mathrm{F}^\mathrm{I} \right] .
\end{equation}
From \eq{gaugeinvtometricEZ}, the even-parity metric perturbation components $\mathrm{A}$, $\mathrm{D}$, $\mathrm{H}$, $\mathrm{K}$ could be read out directly.

\subsubsection{\textbf{Even-parity Perturbations and the RW gauge}}

Similar to the development provided in the previous subsection, using the field equations $E_\mathrm{A}^\mathrm{II}$ to $E_\mathrm{K}^\mathrm{II}$ in Appendix C, one can eliminate the gauge invariants $\delta^\mathrm{II}$ and $\epsilon^\mathrm{II}$, and then obtain the following coupled equations
\begin{align}
\label{prchiRW}
\frac{\pp }{\pp r}\chi^\mathrm{II} =& \frac{2e^{2\Lambda-2\Phi}r}{\tau^\mathrm{II}\eta^{\mathrm{II}}}
\left(\sigma^\mathrm{II}\frac{\pp^2}{\pp t^2}\chi^\mathrm{II}-2\frac{\pp^2}{\pp t^2}\psi^\mathrm{II}
+\frac{1}{2}e^{2\Lambda}r\frac{\pp}{\pp t}E_\mathrm{D}^\mathrm{II}\right)
+\frac{\gamma^\mathrm{II}}{r \tau^\mathrm{II}}\chi^\mathrm{II}+\frac{\rho^\mathrm{II}}{r \tau^\mathrm{II}}\psi^\mathrm{II} \non
& -\frac{e^{2\Lambda}r}{2\tau^\mathrm{II}}(e^{2\Lambda}\lambda+2r\Phi'-2)E_\mathrm{F}^\mathrm{II}
+\frac{e^{2\Lambda}r}{2\tau^\mathrm{II}}\left(2E_\mathrm{H}^\mathrm{II}+E_\mathrm{K}^\mathrm{II}\right) ,
\end{align}
\begin{align}
\label{prpsiRW}
\frac{\pp }{\pp r}\psi^\mathrm{II} =& \frac{e^{2\Lambda-2\Phi}r}{\tau^\mathrm{II}}
\left(\sigma^\mathrm{II}\frac{\pp^2}{\pp t^2}\chi^\mathrm{II}-2\frac{\pp^2}{\pp t^2}\psi^\mathrm{II}
+\frac{1}{2}e^{2\Lambda}r\frac{\pp }{\pp t}E_\mathrm{D}^\mathrm{II}\right)
+\frac{\mu^\mathrm{II}}{r \tau^\mathrm{II}}\chi^\mathrm{II}+\frac{\nu^\mathrm{II}}{r \tau^\mathrm{II}}\psi \non
& +\frac{e^{2\Lambda}r}{4\tau^\mathrm{II}}\left(\kappa^\mathrm{II}E_\mathrm{F}^\mathrm{II}
+2\eta^\mathrm{II}E_\mathrm{H}^\mathrm{II}+\eta^\mathrm{II}E_\mathrm{K}^\mathrm{II}\right)-\frac{1}{4}e^{4\Lambda-2\Phi}rE_\mathrm{A}^\mathrm{II} ,
\end{align}
where the parameters $\sigma^\mathrm{II}$, $\tau^\mathrm{II}$, $\eta^\mathrm{II}$,  $\kappa^\mathrm{II}$, $\rho^\mathrm{II}$, $\mu^\mathrm{II}$, $\nu^\mathrm{II}$ depend only on the background, and the explicit expressions of them can be found in Appendix D.

We find that the master variable can be similarly constructed as
\bean
\label{evenmastervRW}
Z^\mathrm{II}\hsp^{(+)}=\sigma^\mathrm{II}\chi^\mathrm{II}-2\psi^\mathrm{II} ,
\eean
and then the master equation under the RW gauge is given by
\begin{equation}
\label{evenmasterRW}
\left[\frac{2e^{2\Lambda-2\Phi}r}{N^{\mathrm{II}}\eta^{\mathrm{II}}}
\left(4r^2(\Lambda'+\Phi')\sigma^{\mathrm{II}}\frac{\pp^3}{\pp t^2 \pp r}
+\frac{N_\mathrm{T}^\mathrm{II}}{N^\mathrm{II}\eta^{\mathrm{II}}\tau^\mathrm{II}}\frac{\pp^2}{\pp t^2}\right)
-\frac{2r\sigma^{\mathrm{II}}\tau^\mathrm{II}}{N^{\mathrm{II}}}\frac{\pp^2}{\pp r^2}+\frac{N_\mathrm{R}^\mathrm{II}}{\left({N^\mathrm{II}}\right)^2}\frac{\pp}{\pp r}
+\frac{N_\mathrm{Z}^\mathrm{II}}{r\left({N^\mathrm{II}}\right)^2\tau^\mathrm{II}}\right]Z^\mathrm{II}\hsp^{(+)}
=S_{\text{even}}^\mathrm{II},
\end{equation}
with the source term given by
\begin{align}
\label{evensourceRW}
S_{\text{even}}^\mathrm{II} =& \frac{e^{4\Lambda-2\Phi}r}{N^{\mathrm{II}}} \left[\frac{N_\mathrm{A}^\mathrm{II}}{2N^\mathrm{II}}E_\mathrm{A}^\mathrm{II}
-r\sigma^{\mathrm{II}}\tau^\mathrm{II}\frac{\pp}{\pp r}E_\mathrm{A}^\mathrm{II}
+\frac{rN_\mathrm{D}^\mathrm{II}}{N^\mathrm{II}\left(\eta^{\mathrm{II}}\right)^2\tau^\mathrm{II}}\frac{\pp}{\pp t}E_\mathrm{D}^\mathrm{II}
-\frac{4r^3(\Lambda'+\Phi')\sigma^{\mathrm{II}}}{\eta^{\mathrm{II}}}\frac{\pp^2}{\pp t\pp r}E_\mathrm{D}^\mathrm{II} \right]
-\frac{2e^{2\Lambda}r^2 M^{\mathrm{II}}_\mathrm{F}\sigma^{\mathrm{II}}}{N^{\mathrm{II}}}\frac{\pp}{\pp r}
E^{\mathrm{II}}_{\mathrm{F}}\non
& +\frac{e^{2\Lambda}r}{\left(N^{\mathrm{II}}\right)^2\tau^{\mathrm{II}}} \left[\frac{1}{2}N_\mathrm{F}^\mathrm{II}E_\mathrm{F}^\mathrm{II}
+N_\mathrm{HK}^\mathrm{II}\left(E_\mathrm{H}^\mathrm{II}+\frac{1}{2}E_\mathrm{K}^\mathrm{II}\right) \right]-\frac{2e^{2\Lambda}r^3\left(\Lambda
'+\Phi'\right)\sigma^{\mathrm{II}}}{N^{\mathrm{II}}}\left(2\frac{\pp}{\pp r}E^{\mathrm{II}}_{\mathrm{H}}+\frac{\pp}{\pp r}E^{\mathrm{II}}_{\mathrm{K}}\right) ,
\end{align}
where $N^\mathrm{II}$, $N_\mathrm{T}^\mathrm{II}$, $N_\mathrm{R}^\mathrm{II}$, $N_\mathrm{Z}^\mathrm{II}$, $N_\mathrm{A}^\mathrm{II}$, $N_\mathrm{D}^\mathrm{II}$, $M^{\mathrm{II}}_\mathrm{F}$, $N_\mathrm{F}^\mathrm{II}$ all depend on the background, which can be found in Appendix D. Similarly to the case under the EZ gauge, the master equation is a third-order equation, and which can be transformed as a second-order differential equation. Note that when the background metric takes the form
\begin{equation}
ds^2=-f(r)dt^2+f(r)^{-1}dr^2+r^2d\Omega^2
\end{equation}
The relation $\Lambda'+\Phi'=0$ makes the third-order terms in \eq{evenmasterRW} vanishes. In the Schwarzschild case, the master equation \eq{evenmasterRW} will also reduce to the result of Zerilli, see Appendix E. Once \eq{evenmasterRW} is solved, one can get $\chi^\mathrm{II}$ and $\psi^\mathrm{II}$ from
\begin{align}
\chi^\mathrm{II} =& \frac{1}{N^\mathrm{II}}
\left[\frac{8e^{2\Lambda-2\Phi}r^3\left(\Lambda'+\Phi'\right)}
{\eta^\mathrm{II}}\frac{\pp^2}{\pp t^2}Z^\mathrm{II}\hsp^{(+)}
-2r\tau^\mathrm{II}\frac{\pp }{\pp r}Z^\mathrm{II}\hsp^{(+)} +M_1^\mathrm{II}Z^\mathrm{II}\hsp^{(+)} +e^{4\Lambda-2\Phi}r^2\tau^\mathrm{II} E_\mathrm{A}^\mathrm{II} \right. \non
& \left. +\frac{4e^{4\Lambda-2\Phi}r^4\left(\Lambda'+\Phi'\right)}
{\eta^\mathrm{II}}\frac{\pp}{\pp t}E_\mathrm{D}^\mathrm{II}
+2e^{2\Lambda}r^2M^\mathrm{II}_{\mathrm{F}} E_\mathrm{F}^\mathrm{II}
+2e^{2\Lambda}r^3\left(\Lambda'+\Phi'\right)\left(E_\mathrm{K}^\mathrm{II}+2E_\mathrm{H}^\mathrm{II}\right) \right] ,
\end{align}
\begin{align}
\psi^\mathrm{II} =& \frac{\sigma^\mathrm{II}}{2N^\mathrm{II}}
\left[\frac{8e^{2\Lambda-2\Phi}r^3\left(\Lambda'+\Phi'\right)}
{\eta^\mathrm{II}}\frac{\pp^2}{\pp t^2}Z^\mathrm{II}\hsp^{(+)}
-2r\tau^\mathrm{II}\frac{\pp }{\pp r}Z^\mathrm{II}\hsp^{(+)} +\frac{M_2^\mathrm{II}}{\sigma^{\mathrm{II}}}Z^\mathrm{II}\hsp^{(+)} +e^{4\Lambda-2\Phi}r^2\tau^\mathrm{II} E_\mathrm{A}^\mathrm{II} \right. \non
& \left. +\frac{4e^{4\Lambda-2\Phi}r^4\left(\Lambda'+\Phi'\right)}
{\eta^\mathrm{II}}\frac{\pp}{\pp t}E_\mathrm{D}^\mathrm{II}
+2e^{2\Lambda}r^2M^\mathrm{II}_{\mathrm{F}} E_\mathrm{F}^\mathrm{II}
+2e^{2\Lambda}r^3\left(\Lambda'+\Phi'\right)\left(E_\mathrm{K}^\mathrm{II}+2E_\mathrm{H}^\mathrm{II}\right) \right],
\end{align}
and the remaining two gauge invariants $\delta^\mathrm{II}$ and $\epsilon^\mathrm{II}$ can be obtained from $E_\mathrm{D}^\mathrm{II}$ and $E_\mathrm{F}^\mathrm{II}$, which are given respectively by
\begin{equation}
\delta^\mathrm{II} = \frac{1}
{-2-e^{2\Lambda}\left(\lambda-2\right)+4r\Lambda'}\left[e^{2\Lambda}r^2E^{\mathrm{II}}_\mathrm{D}+r\left(2+e^{2\Lambda}\left(\lambda-2\right)+4r\Phi'\right)\frac{\pp}{\pp t}\chi^\mathrm{II}-4r\frac{\pp }{\pp t}\psi^\mathrm{II}\right] ,
\end{equation}
\begin{equation}
\epsilon^\mathrm{II} =e^{-2\Lambda+2\Phi}\left[\left(1-r\Lambda'+r\Phi'\right)\chi^\mathrm{II}
-\psi^\mathrm{II}+r\frac{\pp }{\pp r}\chi^\mathrm{II} +\frac{1}{2}e^{2\Lambda}r^2E_\mathrm{F}^\mathrm{II} \right] .
\end{equation}
When $\chi^\mathrm{II}$, $\psi^\mathrm{II}$, $\delta^\mathrm{II}$ and $\epsilon^\mathrm{II}$ are solved, the even-parity metric perturbation components $\mathrm{A}$, $\mathrm{D}$, $\mathrm{E}$, $\mathrm{K}$ are given by
\begin{align}
\mathrm{A} &= 2e^{-2\Lambda+2\Phi}r\Phi'\chi^\mathrm{II}-2\epsilon^\mathrm{II}, & \mathrm{D} &= \delta^\mathrm{II} +r\frac{\pp }{\pp t}\chi^\mathrm{II}, \non
\mathrm{E} &= -2e^{-2\Lambda}\chi^\mathrm{II}, & \mathrm{K} &= 2\psi^\mathrm{II}+2(r\Lambda'-1)\chi^\mathrm{II}-2r\frac{\pp }{\pp r}\chi^\mathrm{II}.
\end{align}

Under the RW gauge, the master variable can be written as the combination of the gauge invariants $\chi^\mathrm{II}$ and $\psi^\mathrm{II}$. Note that Zerilli constructed the master variable as the combination of $K_{LM}$ and $R_{LM}$ \cite{Zerilli1970PRL}, which correspond to a combination of $\chi^\mathrm{II}$ and $\delta^\mathrm{II}$ in our paper.

\subsubsection{\textbf{Odd-parity Perturbations}}

For the odd-parity perturbations, the EZ gauge and the RW gauge are identical. So, in the following we shall not distinguish them. We construct the odd-parity master variable as
\begin{equation}
\label{oddvariablel2}
Z^{(-)}=\left(r+r^2\Lambda'+r^2\Phi'\right)\beta-r^2\frac{\pp}{\pp r}\beta+r^2\frac{\pp}{\pp t}\alpha ,
\end{equation}
then the master equation is given by
\begin{equation}
\label{oddmasterl2}
\left\{ -\frac{\pp^2}{\pp t^2}+e^{-2\Lambda+2\Phi}
\left[\frac{\pp^2}{\pp r^2}-\left(\frac{X'}{X}+3\Lambda'+3\Phi'\right)\frac{\pp}{\pp r}+\frac{N_{\text{odd}}}{r^2} \right]\right\} Z^{(-)} = S_{\text{odd}} ,
\end{equation}
with the source term given by
\bean
\label{oddsourcel2}
S_{\text{odd}} = e^{2\Phi}r^2\left[\left( \frac{1}{r}+\Lambda'-\Phi'-\frac{X'}{X}\right)E_\mathrm{C}
+\frac{\pp }{\pp r}E_\mathrm{C}+\frac{\pp }{\pp t}E_\mathrm{J}\right] ,
\eean
where $X$ and $N_{\text{odd}}$ are functions only depending on the background, and the explicit expressions of which can be found in Appendix D. Once \eq{oddmasterl2} is solved, together with the structure of the odd-parity master variable \eq{oddvariablel2} and the perturbed field equations \eqs{projECgaugein} and \meq{projEJgaugein}, we have
\begin{equation}
\label{reconsalpha}
\alpha = -\frac{e^{2\Lambda-2\Phi}}{X} \left(e^{2\Phi}r^2E_\mathrm{J}+\frac{\pp }{\pp t}Z^{(-)} \right),
\end{equation}
and
\begin{equation}
\label{reconsbeta}
\beta = \frac{1}{X}\left[ e^{2\Lambda}r^2E_\mathrm{C}-\left(\frac{1}{r}-2\Lambda'-2\Phi' \right)Z^{(-)}-\frac{\pp }{\pp r}Z^{(-)} \right].
\end{equation}
Then, the odd-parity metric perturbation components $\mathrm{C}$ and $\mathrm{J}$ can be read off from these expressions.

\subsection{Specific Cases for \texorpdfstring{$l=0,1$}.}

In the previous subsection, we have discussed the construction of master variables and master equations for even-parity and odd-parity perturbations for the $l\geq 2$ cases. However, the decomposition of the metric perturbation would take some other forms for the specific cases of $l=0, 1$. In this section, we investigate how to solve the metric perturbation for $l=0,1$. Note that in this subsection, we use, e.g. $\mathrm{A}_0$ and $\mathrm{A}_1$, to represent the scalar functions in the metric perturbation and the gauge vector for $l=0$ and $l=1$ cases, respectively. And we also use, e.g. $E_{\mathrm{A}}^{(l=0)}$ and $E_{\mathrm{A}}^{(l=1)}$ to represent the projection function $E_{\mathrm{A}}$ for $l=0$ and $l=1$ cases, respectively.

\subsubsection{\textbf{\texorpdfstring{$l=0$}. case}}

For the special case $l=0$, there remains one scalar spherical harmonic function $Y^{00}=\frac{1}{2\sqrt{\pi}}$, which leads the metric perturbation to
\begin{equation}
h^{(l=0)}_{ab} = \frac{1}{2\sqrt{\pi}}(\mathrm{A}_0v_av_b+2\mathrm{D}_0v_{(a}n_{b)}+\mathrm{E}_0\sigma_{ab}+\mathrm{K}_0n_ab_b),
\end{equation}
and the gauge vector takes the form
\begin{equation}
\xi^{(l=0)}_a = \frac{1}{2\sqrt{\pi}}(\mathrm{P}_0v_a+\mathrm{R}_0n_a).
\end{equation}
In the $l=0$ case, we only have four metric perturbation components, and all of them are even-parity. Under the gauge transformation, the metric perturbation will be transformed as
\begin{align}
\label{DeltaAKl0}
\Delta \mathrm{A}_0 &= -2\frac{\pp }{\pp t}\mathrm{P}_0-2e^{-2\Lambda+2\Phi}\Phi'\mathrm{R}_0,
& \Delta \mathrm{D}_0 &= \left(\frac{\pp }{\pp r}-2\Phi' \right) \mathrm{P}_0 -\frac{\pp }{\pp t}\mathrm{R}_0, \non
\Delta \mathrm{E}_0 &= 2\frac{e^{-2\Lambda}}{r}\mathrm{R}_0,
& \Delta \mathrm{K}_0 &= \left(2\frac{\pp }{\pp r}-2\Lambda' \right)\mathrm{R}_0 .
\end{align}
The structure of $\Delta \mathrm{A}_0$, $\Delta \mathrm{D}_0$ and $\Delta \mathrm{K}_0$ are the same as the $l\geq 2$ cases, but $\Delta \mathrm{E}_0$ is different from the expression given in \eq{DeltaAK} since $\mathrm{S}$ does not exist for $l=0$. From \eq{DeltaAKl0}, we can construct two gauge invariants
\begin{align}
\psi_0 &= \frac{1}{2}\mathrm{K}_0-\frac{r}{2}e^{2\Lambda}\Lambda'\mathrm{E}_0-\frac{1}{2}e^{2\Lambda}\mathrm{E}_0-\frac{r}{2}e^{2\Lambda}\frac{\pp}{\pp r}\mathrm{E}_0 , \non
o_0 &= \frac{1}{2}\frac{\pp }{\pp r}\mathrm{A}_0-\Phi'\mathrm{A}_0+\frac{\pp }{\pp t}\mathrm{D}_0+\frac{1}{2}e^{2\Phi}(\Phi'+r\Phi'')\mathrm{E}_0+\frac{1}{2}e^{2\Phi}r\Phi'\frac{\pp}{\pp r}\mathrm{E}_0
+\frac{1}{2}e^{2\Lambda}r\frac{\pp^2}{\pp t^2}\mathrm{E}_0 .
\end{align}

Choosing the gauge $\mathrm{D}_0=\mathrm{E}_0=0$, we find that $E_\mathrm{A}^{(l=0)}$, $E_\mathrm{D}^{(l=0)}$, $E_\mathrm{E}^{(l=0)}$ and $E_\mathrm{K}^{(l=0)}$ are given by
\begin{align}
E_\mathrm{A}^{(l=0)} =& 2r^{-2}e^{-2\Lambda}(-1+e^{2\Lambda}+2r\Lambda')\mathrm{A}_0  +2r^{-2}e^{-4\Lambda+2\Phi}(4r\Lambda'-1)\mathrm{K}_0-2r^{-1}e^{-4\Lambda+2\Phi}\frac{\pp }{\pp r}\mathrm{K}_0 , \non
E_\mathrm{D}^{(l=0)} =& 2r^{-1}e^{-2\Lambda}\frac{\pp }{\pp t}\mathrm{K}_0 , \non
E_\mathrm{E}^{(l=0)} =& 2r^{-1}e^{-2\Lambda-2\Phi}(r\Lambda'\Phi'-r\Phi''-\Phi')\mathrm{A}_0-r^{-1}e^{-2\Lambda-2\Phi}(r\Lambda'+2r\Phi'-1)\frac{\pp }{\pp r}\mathrm{A}_0+e^{-2\Lambda-2\Phi}\frac{\pp^2}{\pp r^2}\mathrm{A}_0 \non
& +2r^{-1}e^{-4\Lambda}(\Phi'+r{\Phi'}^2+r\Phi''-2r\Lambda'\Phi'-2\Lambda')\mathrm{K}_0+r^{-1}e^{-4\Lambda}(r\Phi'+1)\frac{\pp }{\pp r}\mathrm{K}_0
+e^{-2\Lambda-2\Phi}\frac{\pp^2}{\pp t^2}\mathrm{K}_0 , \non
E_\mathrm{K}^{(l=0)} =& -4r^{-1}e^{-2\Phi}\Phi'\mathrm{A}_0+2r^{-1}e^{-2\Phi}\frac{\pp }{\pp r}\mathrm{A}_0
+2r^{-2}\mathrm{K}_0 .
\end{align}
The gauge invariants $\psi_0$ and $o_0$ can be constructed as
\begin{equation}
\label{psil0}
\psi_0 = \frac{1}{2}\mathrm{K}_0, \quad\quad o_0= \frac{1}{2}\frac{\pp }{\pp r}\mathrm{A}_0-\Phi'\mathrm{A}_0.
\end{equation}
And then
$E_\mathrm{D}^{(l=0)}$, $E_\mathrm{E}^{(l=0)}$ and $E_\mathrm{K}^{(l=0)}$ can be written as
\begin{align}
E_\mathrm{D}^{(l=0)} =& 4r^{-1}e^{-2\Lambda}\frac{\pp}{\pp t}\psi_0, \label{EDl0} \\
E_\mathrm{E}^{(l=0)} =& -2r^{-1}e^{-2\Lambda-2\Phi}(r\Lambda'-1)o_0+2e^{-2\Lambda-2\Phi}\frac{\pp}{\pp r}o_0+4r^{-1}e^{-4\Lambda}(\Phi'-2\Lambda'+r{\Phi'}^2+r\Phi''-2r\Lambda'\Phi')\psi_0 \non
& +2r^{-1}e^{-4\Lambda}(r\Phi'+1)\frac{\pp }{\pp r}\psi_0+2e^{-2\Lambda-2\Phi}\frac{\pp^2}{\pp t^2}\psi_0, \label{EEl0}  \\
E_\mathrm{K}^{(l=0)} =& 4r^{-1}e^{-2\Phi}o_0+4r^{-2}\psi_0. \label{EKl0}
\end{align}

From \eq{EKl0}, we have
\begin{equation}
\label{o0l0}
o_0=\frac{1}{4}e^{2\Phi}\left(rE_\mathrm{K}^{(l=0)}-\frac{4}{r}\psi_0\right) .
\end{equation}
Substituting \eqs{EDl0} and \meq{o0l0} into \eq{EEl0}, we find that $\psi_0$ satisfied
\begin{equation}
\label{psi02}
\left(\iota_0 \frac{\pp }{\pp r}-\sigma_0\right)\psi_0=\frac{r}{2}e^{4\Lambda}S^{(l=0)},
\end{equation}
where $\iota_0$ and $\sigma_0$ are functions depend on background, which can be found in Appendix D, and $S^{(l=0)}$ is the source term,
\begin{equation}
S^{(l=0)}=\frac{1}{2}e^{-2\Phi}r\frac{\pp }{\pp t}E^{(l=0)}_{\mathrm{D}}-E^{(l=0)}_{\mathrm{E}}
+\frac{1}{2}e^{-2\Lambda}\left[r\frac{\pp }{\pp r}E_{\mathrm{K}}^{(l=0)}+\left(2-r\Lambda'+2r\Phi'\right)E_{\mathrm{K}}^{(l=0)}\right].
\end{equation}

\subsubsection{\textbf{\texorpdfstring{$l=1$}. case}}

For the case $l=1$, the tensor harmonic basis $T_{ab}^{E2,1m}$ and $T_{ab}^{B2,1m}$ vanish, hence the scalar functions $\mathrm{F}_1$ and $\mathrm{G}_1$ would no longer exist. The four components of the gauge vector $\xi_a$ imply that there are only three gauge invariants for even-parity and one gauge invariant for odd-parity. The remaining eight projections of $2\grad_{(a}\xi_{b)}$ are the same as for $l\geq2$ cases.

First, we investigate the even-parity sector. As we explain for $l\geq2$ cases, one should first determine the function $\mathrm{S}$ of the gauge vector $\xi_a$ to make $\mathrm{F}$ vanish under the gauge transformation. However, when $l=1$, the lack of $\mathrm{F}_1$ prevents us from constructing the gauge invariants as the cases for $l\geq2$. The even-parity $l=1$ metric perturbations are related to the linear momentum of the system \cite{Zerilli1970PRD}.
Note that if we take the gauge choice $\mathrm{B}_1=\mathrm{E}_1=\mathrm{H}_1=0$, which was introduced by Zerilli in the Schwarzschild spacetime, the relation between $E_\mathrm{A}$ and metric components would no longer be a simple relationship. However, taking the gauge
\begin{equation}
\mathrm{A}_1=\mathrm{E}_1=\mathrm{H}_1=0,
\end{equation}
we find that $E_\mathrm{A}^{(l=1)}$ and the metric perturbed component $\mathrm{K_1}$ have a simple relation. In particular, we find
\begin{align}
E_\mathrm{A}^{(l=1)} =& -2r^{-1}e^{-4\Lambda+2\Phi}\left(\frac{\pp }{\pp r}\mathrm{K}_1+r^{-1}(1+e^{2\Lambda}-4r\Lambda')\mathrm{K}_1 \right) , \label{EAl1} \\
E_\mathrm{D}^{(l=1)}  =& 2r^{-2}(1-2r\Phi')\mathrm{B}_1+2r^{-1}\frac{\pp}{\pp r}\mathrm{B}_1+ 2r^{-1}e^{-2\Lambda} \frac{\pp }{\pp t}\mathrm{K}_1-2r^{-2}e^{-2\Lambda}(1-2r\Lambda')\mathrm{D}_1, \label{EDl1} \\
E_\mathrm{K}^{(l=1)} +4E_\mathrm{H}^{(l=1)}
=& 2r^{-2}e^{-2\Lambda}(-2+e^{2\Lambda}-2r\Phi')\mathrm{K}_1  +4r^{-1}e^{-2\Phi}(1-e^{2\Lambda})\frac{\pp}{\pp t}\mathrm{B}_1-4e^{-2\Phi}\frac{\pp^2}{\pp t\pp r}\mathrm{B}_1. \label{EKl1}
\end{align}
From \eq{EAl1}, the metric perturbation function $\mathrm{K}_1$ can be found. This solution can be used in \eq{EKl1} to solve $\mathrm{B}_1$, and then from \eq{EDl1}, $\mathrm{D}_1$ could also be solved. The remaining quantities, such as $E_\mathrm{B}$ or $E_\mathrm{E}$, can be used to check the consistency of the solutions.

Then we investigate the odd-parity sector. For $l\geq2$ cases, one should determine the function $\mathrm{Q}$ of the gauge vector $\xi_a$ to make $\mathrm{G}$ vanish under the gauge transformation. For $l=1$ case, $\mathrm{G}_1$ no longer exists, which means that we can not construct the gauge invariants $\alpha$ and $\beta$ as in the $l\geq 2$ cases. However, from the components of the projection of $2\grad_{(a}\xi_{b)}$
\begin{align}
\Delta \mathrm{C}_1 =& -\frac{\pp }{\pp t}\mathrm{Q},\\
\Delta \mathrm{J}_1 =& \left(\frac{\pp}{\pp r}-\frac{1}{r}-\Phi'-\Lambda'\right)\mathrm{Q},
\end{align}
we can construct a gauge invariant property $\alpha_1$ as
\begin{equation}
\label{mastervoddl1}
\alpha_1 =r^2\frac{\pp }{\pp t}\mathrm{J}_1+r^2\frac{\pp }{\pp r}\mathrm{C}_1-r\left(1+r\Phi'+r\Lambda' \right)\mathrm{C}_1 .
\end{equation}
Then, for $l=1$, the odd-parity of the projection of EFEs are
\begin{align}
E_\mathrm{C}^{(l=1)}  =
& -r^{-1}e^{-2\Lambda}(-2+3r\Lambda'+3r\Phi')\frac{\pp }{\pp r}\mathrm{C}_1+e^{-2\Lambda}\frac{\pp^2 }{\pp r^2}\mathrm{C}_1
-r^{-1}e^{-2\Lambda}(-3+2r\Lambda'+2r\Phi')\frac{\pp }{\pp t}\mathrm{J}_1+e^{-2\Lambda}\frac{\pp^2 }{\pp t\pp r}\mathrm{J}_1 , \label{ECl1}\non
&-r^{-2}e^{-2\Lambda}\left(
-2+2r^2{\Lambda'}^2-3r\Phi'+r\Lambda\left(1+6r\Phi' \right)-r^2\Lambda''-3r^2\Phi''\right)\mathrm{C}_1 \\
E_\mathrm{J}^{(l=1)}  =& -r^{-1}e^{-2\Lambda}\left(
2\Lambda'\left(1+r\Phi'\right)-2(\Phi'+r{\Phi'}^2+r\Phi'')
\right)\mathrm{J}_1  -e^{-2\Phi}\frac{\pp^2}{\pp t^2}\mathrm{J}_1\non
&+ r^{-1}e^{-2\Phi}(1+r\Lambda'+r\Phi')\frac{\pp }{\pp t}\mathrm{C}_1-e^{-2\Phi}\frac{\pp^2}{\pp t\pp r}\mathrm{C}_1 \label{EJl1}.
\end{align}

Together with \eq{mastervoddl1}, \eq{ECl1} and \eq{EJl1} can be decoupled, and the master equation for the odd-parity perturbation is given by
\begin{equation}
\label{mastereqoddl1}
\left\{ -\frac{\pp^2}{\pp t^2}+e^{-2\Lambda+2\Phi}
\left[\frac{\pp^2}{\pp r^2}-\left(\frac{X_1'}{X_1}+3\Lambda'+3\Phi'\right)\frac{\pp}{\pp r}
+\frac{N_{\text{odd}}^{(l=1)}}{r^2} \right] \right\} \alpha_1=S_{\text{odd}}^{(l=1)} ,
\end{equation}
where $S_{\text{odd}}^{(l=1)}$ is the source term,
\begin{equation}
S_{\text{odd}}^{(l=1)}=e^{2\Phi}r^2\left[\left(\frac{1}{r}+\Lambda'-\Phi'-\frac{X_1'}{X_1}\right)E_\mathrm{C}^{(l=1)}
+\frac{\pp}{\pp r}E_\mathrm{C}^{(l=1)}  +\frac{\pp}{\pp t}E_\mathrm{J}^{(l=1)}   \right].
\end{equation}
In the above equations, $X_1$ and $N_{\text{odd}}^{(l=1)}$ are all depend only on the background, which can be found in Appendix D. Once \eq{mastereqoddl1} is solved, the metric perturbation components $\mathrm{C}_1$ and $\mathrm{J}_1$ can be determined by
\begin{align}
\mathrm{C}_1 =& -\frac{1}{X_1}\left[e^{2\Lambda}r^2E_\mathrm{C}^{(l=1)} -\left(\frac{1}{r}-2\Lambda'-2\Phi'\right)\alpha_1-\frac{\pp }{\pp r}\alpha_1 \right], \label{C1} \\
\mathrm{J}_1 =& -\frac{e^{2\Lambda-2\Phi}}{X_1}\left(e^{2\Phi}r^2E_\mathrm{J}^{(l=1)} +\frac{\pp }{\pp t}\alpha_1 \right). \label{J1}
\end{align}

\section{A Point Particle as the Source}

In this section, we present a simple example that a small object moves along a circular orbit around the center of a spherically symmetric spacetime. We analyse the solutions for $l=0,1$ in this section. For $l\geq 2$ cases, we only provide a general outline.

Assuming that the small object moves along the worldline $z(\tau)$ with mass $\mu$ and four velocity $u_a$, the stress-energy tensor of this point particle takes the form \cite{Poisson2011}
\begin{equation}
T_{ab}=\int\frac{\mu u_au_b}{\sqrt{-g^{(0)}}}\delta^4[x-z(\tau)]d\tau ,
\end{equation}
where $\tau$ is the proper time, $g^{(0)}$ and $\delta^4$ are the determinant of the background and the four-dimensional Dirac delta function, respectively. Generalizing the standard analysis from the textbook \cite{Waldbook}, the time-like geodesics in general spherically symmetric spacetime is
\begin{equation}
-1 =  g_{ab}u^au^b = -e^{2\Phi}{\dot t}^2+e^{2\Lambda}{\dot r}^2+r^2{\dot\varphi}^2.
\end{equation}
Using the static Killing field $\xi^a=(\pp/\pp t)^a$ and the rotational Killing field $\psi^a=(\pp/\pp\varphi)^a$, two conserved quantities $\mathcal{E}$ and $L$ can be defined as
\begin{equation}
\mathcal{E} = -g_{ab}\xi^a u^b = e^{2\Phi}\dot t, ~~~~
L = g_{ab}\psi^au^b = r^2 \dot\varphi .
\end{equation}
Then the geodesic equation reads
\begin{equation}
\frac{1}{2}{\dot r}^2+\frac{1}{2}e^{-2\Lambda}\left(\frac{L^2}{r^2}+1\right)=\frac{1}{2}e^{-2\Lambda-2\Phi}\mathcal{E}^2 .
\end{equation}

Considering that the particle moves along a circular orbit with radio $R$, which is determined by $\pp V/\pp r=0$, the four velocity of the particle can be written as
\begin{equation}
u_a= (-\mathcal{E},0,0,L),
\end{equation}
and now $\mathcal{E}$ and $L$ represent the energy and angular momentum of the particle, given respectively by,
\begin{equation}
\label{exampleEL}
\mathcal{E} = \frac{e^{\Phi|_R}}{\sqrt{R\Lambda'|_R+1}},  ~~~~ L = R\sqrt{\frac{-R\Lambda'|_R}{R\Lambda'|_R+1}}.
\end{equation}
where $|_R$ denotes the corresponding function taking its value along the orbit. The orbital frequency can also be defined as
\begin{equation}
\Omega^2=\left(\frac{u^\varphi}{u^t}\right)^2 = -\frac{e^{2\Phi}\Lambda'|_R}{R} ,
\end{equation}
which gives two useful relations
\begin{equation}
\label{energyangular}
\mathcal{E} = \frac{e^{2\Phi|_R}}{\Omega R^2}L, ~~~~~~~~ \Omega L = -R\Lambda'|_R\cdot\mathcal{E}.
\end{equation}

Projecting the stress-energy tensor $T_{ab}$ to the harmonic basis, one can obtain $E_\mathrm{A}$-$E_\mathrm{K}$. The results reveal that $E_\mathrm{D}$, $E_\mathrm{H}$, $E_\mathrm{J}$ and $E_\mathrm{K}$ vanish automatically, and the non-vanishing projections of the stress-energy tensor are given as follows. For $l\geq 0$, we have
\begin{align}
E_\mathrm{A}^{lm} =& -16\pi e^{-\Lambda+\Phi}\frac{\mu \mathcal{E}}{r^2}\delta(r-R)Y_{lm}^*\left(\frac{\pi}{2},\Omega t\right),  \label{exampleEA}\\
E_\mathrm{E}^{lm} =& -8\pi e^{-\Lambda-\Phi}\frac{\mu L\Omega R^2}{r^4}\delta (r-R)Y_{lm}^*\left(\frac{\pi}{2},\Omega t\right), \label{exampleEE}
\end{align}
where $E_\mathrm{A}$ and $E_\mathrm{E}$ satisfy the relation $E_\mathrm{A}=-\frac{2e^{2\Phi}r^2}{R^3\Lambda'}E_\mathrm{E}$. For $l\geq 1$, we find
\begin{align}
E_\mathrm{B}^{lm} =& -\frac{16\pi}{l(l+1)}e^{-\Lambda-\Phi}\frac{\mu \mathcal{E}\Omega R^2}{r^3}\delta (r-R)
\left(\frac{\pp }{\pp \varphi}Y_{lm}^*(\theta,\varphi)\right)\Big|_{\theta=\frac{\pi}{2},\varphi=\Omega t}, \label{exampleEB}\\
E_\mathrm{C}^{lm} =& -\frac{16\pi}{l(l+1)}\frac{\mu \mathcal{E} \Omega R^2}{r^3}\delta(r-R)
\left(\frac{\pp }{\pp \theta}Y^*_{lm}(\theta,\varphi)\right)\Big|_{\theta=\frac{\pi}{2},\varphi=\Omega t} , \label{exampleEC}
\end{align}
and for $l\geq 2$,
\begin{align}
E_\mathrm{F}^{lm} =& \frac{16\pi(l-2)!}{(l+2)!}e^{-\Lambda-\Phi}\frac{\mu L \Omega R^2}{r^4}\delta(r-R)
\left[2\frac{\pp^2}{\pp \theta^2} Y^*_{lm}(\theta,\varphi) +l(l+1)Y_{lm}^*(\theta,\varphi)\right]\Big|_{\theta=\frac{\pi}{2},\varphi=\Omega t}, \label{exampleEF} \\
E_\mathrm{G}^{lm} =& -\frac{32\pi (l-2)!}{(l+2)!}\frac{\mu L \Omega R^2}{r^4}\delta (r-R)
\left(\frac{\pp^2}{\pp \theta \pp \varphi}Y^*_{l,m}(\theta,\varphi)\right)\Big|_{\theta=\frac{\pi}{2},\varphi=\Omega t}. \label{exampleEG}
\end{align}

\subsection{Perturbations for \texorpdfstring{$l=0$}.}

For $l=0$, substituting \eqs{exampleEE} into \eq{psi02}, we have
\begin{equation}
\left(\iota_0 \frac{\pp }{\pp r}-\sigma_0\right)\psi_0= -\frac{r}{2}e^{4\Lambda}E_\mathrm{E}^{00}.
\end{equation}
Using the standard method to solve the above equation, one can obtain
\begin{align}
\psi_0 =& -\exp\left(\int^r\frac{\sigma_0(r)}{\iota_0(r)}dr\right)
\int^{r}\left[\frac{r'e^{4\Lambda(r')}}{2\iota_0(r')}\text{exp}\left(-\int^{r'}\frac{\sigma_0(r'')}{\iota_0(r'')}dr'' \right) E^{00}_\mathrm{E}(r') \right]dr'  \non
=& -\exp\left({\int^r\frac{\sigma_0}{\iota_0}dr-\int^R\frac{\sigma_0}{\iota_0}dr }\right)
\frac{R e^{4\Lambda|_R}}{2\iota_0|_R}\bar{E}^{00}_\mathrm{E}\Theta (r-R) \non
=& -\exp\left(\int^r_R\frac{\sigma_0}{\iota_0}dr\right)
\frac{R e^{4\Lambda|_R}}{2\iota_0|_R}\bar{E}^{00}_\mathrm{E}\Theta (r-R) ,
\end{align}
where $\Theta(r-R)$ is the unit step function, and $\bar{E}_\mathrm{E}^{00}$ denote the evaluated coefficient given by
\begin{equation}
\bar{E}^{00}_{\mathrm{E}}=-4\sqrt{\pi}e^{-\Lambda|_R-\Phi|_R}\frac{\mu L \Omega}{R^2}= 4\sqrt{\pi}e^{-\Lambda|_R-\Phi|_R}\frac{\mu \Lambda'|_R \cdot\mathcal{E}}{R}.
\end{equation}
From \eq{psil0}, we have
\begin{equation}
\mathrm{K}_0=2\psi_0=-4\sqrt{\pi}\frac{\Lambda'|_R \cdot \mu\mathcal{E}}{\iota_0|_R}
\exp\left(\int_R^r\frac{\sigma_0}{\iota_0}dr+3\Lambda|_R-\Phi|_R\right)\Theta(r-R).
\end{equation}
Using \eq{o0l0}, we can get $o_0$, and then solve \eq{psil0} to obtain $\mathrm{A}_0$
\begin{equation}
\mathrm{A}_0 = 4\sqrt{\pi}e^{2\Phi}\frac{\Lambda'|_R \cdot \mu \mathcal{E}}{\iota_0|_R}\cdot \mathcal{A}\cdot
\exp\left[-\int^R\frac{\sigma_0}{\iota_0}dr+3\Lambda|_R-\Phi|_R\right]\Theta(r-R) ,
\end{equation}
where
\begin{equation}
\mathcal{A}=\int r^{-1}
\exp\left[ \int^r \frac{\sigma_0(r')}{\iota_0(r')}dr'\right]dr.
\end{equation}

With $\mathrm{A}_0$ and $\mathrm{K}_0$ in hand, the metric perturbation can be directly written out. Since both $\mathrm{A}_0$ and $\mathrm{K}_0$ contain step function, it is obvious that inside the orbit, i.e. $r<R$, the perturbation would vanishes. While outside the orbit, i.e. $r>R$, the perturbed metric can be given by
\begin{align}
h_{ab}^{00} =& \frac{1}{2\sqrt\pi}(\mathrm{A}_0 v_av_b+\mathrm{K}_0n_an_b) \non
=& 2e^{2\Phi}\frac{\Lambda'|_R \cdot \mu\mathcal{E}}{\iota_0|_R} \cdot \mathcal{A}\cdot
\exp\left[ -\int^R\frac{\sigma_0}{\iota_0}dr+3\Lambda|_R-\Phi|_R\right] v_av_b \non
& -2\frac{ \Lambda'|_R \cdot \mu\mathcal{E}}{ \iota_0|_R}
\exp\left[\int^r_R\frac{\sigma_0}{\iota_0}dr+3\Lambda|_R-\Phi|_R \right]n_an_b.
\end{align}
Unlike the Schwarzschild spacetime, there is no clear definition of the mass $M$ for general spherically symmetric spacetimes. It is obvious that the perturbation for $l=0$ only affects $g_{tt}$ and $g_{rr}$. In the Schwarzschild spacetime, this perturbation is equivalent to adding an extra mass $\delta m$ to the system \cite{Zerilli1970PRD}.

\subsection{Perturbations for \texorpdfstring{$l=1$}.}

\subsubsection{\textbf{Even-parity Perturbations}}

For the even-parity perturbations with $l=1$, we find that we can set directly $\mathrm{A}_1=\mathrm{E}_1=\mathrm{H}_1=0$. From \eq{EAl1}, we find that $\mathrm{K}_1$ can be determined by $E_\mathrm{A}$. Note that $Y_{1m}^*(\frac{\pi}{2},\varphi)$ would vanish for $m=0$, hence
\begin{equation}
\mathrm{K}_1= -8\pi R\frac{\mu\mathcal{E}}{r^2}Y_{1,\pm1}^*\left(\frac{\pi}{2},\Omega t\right)
\exp\left({-\int_R^r\frac{1+e^{2\Lambda}-4r\Lambda'}{r}dr+3\Lambda|_R-\Phi|_R}\right)\Theta(r-R).
\end{equation}
Substituting $\mathrm{K}_1$ into \eq{EKl1}, together with $E_\mathrm{H}=E_\mathrm{K}=0$, we have
\begin{equation}
\label{exampleKl1}
\frac{\pp}{\pp r}\left(\frac{\pp}{\pp t}\mathrm{B}_1\right)-r^{-1}(1-e^{2\Lambda})\frac{\pp}{\pp t}\mathrm{B}_1
= \mathcal{K}\mathrm{K}_1 ,
\end{equation}
where
\begin{equation}
\mathcal{K}=\frac{1}{2}r^{-2}e^{2\Phi-2\Lambda}(-2+e^{2\Lambda}-2r\Phi') .
\end{equation}
Hence, $\mathrm{B}_1$ can be solved by \eq{exampleKl1}. Then, considering \eq{EDl1} with $E_\mathrm{D}=0$, in principle, the metric perturbed component $\mathrm{D}_1$ can also be obtained.

\subsubsection{\textbf{Odd-parity Perturbations}}

To study the metric perturbation components $\mathrm{C}_1$ and $\mathrm{J}_1$, we should first solve the master variable $\alpha_1$ from \eq{mastereqoddl1}. Then using \eqs{C1} and \meq{J1}, $\mathrm{C}_1$ and $\mathrm{J}_1$ can be obtained. However, $\mathrm{G}$ automatically vanish for the $l=1$ case, there exists one freedom for the odd-parity. If we impose the gauge $\mathrm{C}_1=0$, the master variable becomes
\begin{equation}
\alpha_1=r^2\frac{\pp}{\pp t}\mathrm{J}_1 .
\end{equation}
Then, \eq{C1} becomes
\begin{equation}
\frac{\pp }{\pp r}\alpha_1+\left(\frac{1}{r}-2\Lambda'-2\Phi' \right)\alpha_1=e^{2\Lambda}r^2E_{\mathrm{C}} .
\end{equation}
The solution of this equation is
\begin{align}
\alpha_1 =& -\frac{16\pi}{l(l+1)}\cdot\frac{\mu\mathcal{E}\Omega R^2}{r} \exp(2\Lambda+2\Phi-2\Phi|_R)
\cdot\left(\frac{\pp }{\pp \theta}Y^*_{lm}(\theta,\varphi)\right)|_{\theta=\frac{\pi}{2},\varphi=\Omega t} \non
=& -8\pi \frac{\mu L}{r}e^{2\Lambda+2\Phi}\left(\frac{\pp }{\pp \theta}Y^*_{lm}(\theta,\varphi)\right)|_{\theta=\frac{\pi}{2},\varphi=\Omega t} ,
\end{align}
where we have usedamour the relation \eq{energyangular} and the fact $l=1$. From $\alpha_1$, one can solve $\mathrm{J}_1$ with a undetermined function that only depend on $r$. However, the metric perturbation component $\mathrm{J}_1$ yields a contribution to $h^{1m}_{r\theta}$ and $h^{1m}_{r\varphi}$.

Next, we impose the gauge $\mathrm{J}_1=0$, and now the gauge invariant $\alpha_1$ becomes
\begin{equation}
\alpha_1=r^2\frac{\pp}{\pp r}\mathrm{C}_1-r\left(1+r\Lambda'+r\Phi'\right)\mathrm{C}_1 .
\end{equation}
Substituting this relation into \eqs{C1} and \meq{J1}, we know that $\alpha_1$ should only be a function of $r$, and $\mathrm{C}_1$ satisfies the second-order differential equation
\begin{align}
& r^2\frac{\pp^2 }{\pp r^2}\mathrm{C}_1+r\left(2-3r\Lambda'-3r\Phi'\right)\frac{\pp }{\pp r}\mathrm{C}_1 - \left(
2-2r^2{\Lambda'}^2+3r\Phi'-r\Lambda'(1+6r\Phi')+r^2\Lambda''+3r^2\Phi''
\right)\mathrm{C}_1 =e^{2\Lambda}r^2E_\mathrm{C} ,
\end{align}
which is \eq{ECl1} with $\mathrm{J}_1=0$. Considering that $E_\mathrm{C}$ contains the delta function, we predict that the solution of $\mathrm{C}_1$ contain the step function $\Theta(r-R)$, with two arbitrary constants of integration. One can be determined by considering that the background spacetime without
intrinsic angular momentum, and the other can be determined by the fact that perturbation $\mathrm{C}_1$ would be convergent at infinity \cite{Thompson2016}. If the background metric degenerates to the Schwarzschild spacetime, our equation would become Eq. (10.21) of \cite{Thompson2016}, which gives a non-vanishing $h_{t\varphi}^{01}$ outside the circular orbit to describe the adding angular momentum of the system.

\subsection{Perturbations For \texorpdfstring{$l\geq 2$}.}

Generally speaking, for $l\geq 2$ the problem is much more mathematically involved, and normally we have to seek help from numerical computations. From the source term, \eqs{evensourceEZ}, \meq{evensourceRW} or \meq{oddsourcel2}, we know that they depend only on \eqs{exampleEA}-\meq{exampleEG}. All these terms are proportional to $e^{-im\Omega t}$, which implies that our master equations, \eqs{evenmasterEZ}, \meq{evenmasterRW} and \meq{oddmasterl2}, can be written as the second-order ordinary differential equations in $r$ for each $l$ and $m$. And the distributional sources of these differential equations would be vanish everywhere except at the circular orbit radius $r=R$.

An basic outline is as follows. To solve the second-order equations, one can study two regions separately. One region is from the circular orbit radius to infinity, i.e. $r\in(R,\infty)$. In this region, a naturally boundary condition is considering an appropriate radiation at spatial infinity. In the other region there are several different situations. For example, if the background spacetime has an event horizon, then one should determine this region from the event horizon to the circular orbit radius, i.e., $r\in (r_+,R)$. Then solve the differential equation with the boundary condition at the event horizon $r_+$. Another example is that the background spacetime is a perfect fluid star without event horizons, then one should determine the other region from the center of the star to the circular orbit radius, i.e., $r\in (0,R)$. Then solve the differential equation with some boundary conditions at the center of the star. Finally, matching the solutions obtained in the two separate regions properly across the boundary $r=R$, we obtain the perturbations valid over the whole spacetime.


\section{Conclusions and Discussions}

In this paper, we systematically study the gauge invariant perturbations of a general spherically symmetric background. First, we find that, in general spherically symmetric spacetimes, for even-parity, there are several gauge choices. One is the well-known RW gauge, and the other is the EZ gauge. For odd-parity perturbation, only one gauge choice exists, that is, by setting $\mathrm{G}=0$. Then, we mainly focus on the construction of master equations for $l\geq 2$. For even-parity perturbation,  under the EZ gauge or the RW gauge, the master equation \meq{evenmasterEZ} or \meq{evenmasterRW} are the third-order equations. However, after the separation of variables, the equations all reduce to a second-order equation. For odd-parity perturbation, the master equation is also constructed as \eq{oddmasterl2}. Next, the cases for $l=0$ and $l=1$ are discussed. For the $l=0$ case, we present the equations that the gauge invariants satisfy. For even-parity perturbations with $l=1$, we find that the metric perturbation components are still determined when the source is specified. And for odd-parity perturbations with $l=1$, the master equation \meq{mastereqoddl1} is still a wave-like equation. Finally, using our general results, we investigate a point particle moving along a circular orbit in general spherically symmetric spacetimes. In particular, the form of the solutions for $l=0$ and $l=1$ are carefully discussed.

Our results can be applied to various modified theories of gravity, in which the background is described by the general static metric (\ref{ABneq1}), instead of
the particular one (\ref{ABeq1}). In particular, it can be applied to  the EOB system. Jing et. al. pointed out that one should consider the  Hamilton equations for an EOB system self-consistently \cite{Jing2022}. Specifically, considering that the EOB system takes the spinless effective metric $g_{\mu\nu}^{eff}$ \cite{Damour1999, Damour2016, Damour2018}, the Hamiltonian $H[g_{\mu\nu}^{eff}]$ and the radiation-reaction force $\mathcal{F}^{circ}_\varphi[g_{\mu\nu}^{eff}]$ should be both based on the same effective metric. Under the quasi-circular approximation, the radiation-reaction force can be obtained by the energy-loss rate,
\begin{equation}
\mathcal{F}^{circ}_\varphi[g_{\mu\nu}^{eff}]\simeq \frac{1}{\dot{\varphi}}\frac{dE[g_{\mu\nu}^{eff}]}{dt}.
\end{equation}
Considering that the perturbed Weyl tensor $\psi_4^B$ can be divided into the even-parity $\psi_4^{BE}$ and the odd-parity $\psi_4^{BO}$ parts \cite{Jing2022}, then the energy-loss rate can be calculated from $\psi_4^{BE}$ and $\psi_4^{BO}$ via the relation,
\begin{equation}
\frac{dE[g_{\mu\nu}^{eff}]}{dt}=\frac{c^3}{16\pi G\omega^2}\int\left\{\left[Re(\psi_4^{BE}+\psi_4^{BO})\right]^2+\left[Im(\psi_4^{BE}+\psi_4^{BO})\right]^2\right\}r^2d\Omega^2.
\end{equation}
So, the key step to obtain a self-consistent radiation-reaction force $\mathcal{F}^{circ}_\varphi[g_{\mu\nu}^{eff}]$ is to solve the solution of $\psi_4^{BE}$ and $\psi_4^{BO}$. In \cite{Jing2022}, the authors constructed the decoupled equations for both even-parity $\psi_4^{BE}$ and odd-parity $\psi_4^{BO}$ in the effective metric spacetime rather than in the Schwarzschild spacetime. Generally speaking, $\psi_4^{BO}$ is only related to the odd-parity perturbation, i.e. the $\mathrm{C}$ and $\mathrm{J}$ terms defined in Eq.(\ref{eq17}). When \eq{oddmasterl2} is solved, one can determine $\mathrm{C}$ and $\mathrm{J}$ by \eqs{reconsalpha} and \meq{reconsbeta}, and then $\psi_4^{BO}$ can be calculated. Similar processing can be done for the even-parity perturbation. However, Ref. \cite{Jing2022} only considered the background effective metric that takes the form as \eq{ABeq1}, which can be applied to the Post-Minkowskian (PM) approximation \cite{Damour2016,Damour2018,He2021}, if the undeterminated parameters $d_i$ are well constrained. While in this paper we consider the most general spherical symmetric metric, which can be applied to the EOB theory with either the PN approximation or the PM approximation. We wish to come back to this important issue soon.

\acknowledgments
We would like very much to thank Xiaokai He and Tao Zhu for valuable discussions, and also thank Zhucun Li. This work was partially supported by the National Key Research and Development Program of China Grant No.2020YFC2201503, the National Natural Science Foundation of China under Grants No. 11705053, No. 11975203 and No. 12035005, and the Hunan Provincial Natural Science Foundation of China under Grant No. 2022JJ40262. The work of Fang was supported in part by China Scholarship Council for the visiting post-doc program at Baylor University.

\appendix
\section{Normalization Factors And A-K Decomposition}
The normalization factors $N^{(vec)}(A,r,l)$ and $N^{(ten)}(A,r,l)$ in Sec.II can be calculated as
\begin{align*}
& \oint Y_a^{E,lm}(Y^a_{E,l'm'})d\Omega=N^{(vec)}(E,r,l)\delta_{ll'}\delta_{mm'}=l(l+1)\delta_{ll'}\delta_{mm'}, \\
& \oint Y_a^{B,lm}(Y^a_{B,l'm'})d\Omega=N^{(vec)}(B,r,l)\delta_{ll'}\delta_{mm'}=e^{-2\Lambda-2\Phi}l(l+1)\delta_{ll'}\delta_{mm'}, \\
& \oint Y_a^{R,lm}(Y^a_{R,l'm'})d\Omega=N^{(vec)}(R,r,l)\delta_{ll'}\delta_{mm'}=e^{-2\Lambda}\delta_{ll'}\delta_{mm'},
\end{align*}
and
\begin{align*}
& \oint T_{ab}^{T0,lm}(T^{ab}_{T0,l'm'})^*d\Omega=N^{(ten)}(T0,r,l)\delta_{ll'}\delta_{mm'}=2\delta_{ll'}\delta_{mm'}, \\
& \oint T_{ab}^{L0,lm}(T^{ab}_{L0,l'm'})^*d\Omega=N^{(ten)}(L0,r,l)\delta_{ll'}\delta_{mm'}=e^{-4\Lambda}\delta_{ll'}\delta_{mm'}, \\
& \oint T_{ab}^{E1,lm}(T^{ab}_{E1,l'm'})^*d\Omega=N^{(ten)}(E1,r,l)\delta_{ll'}\delta_{mm'}=\frac{1}{2}e^{-2\Lambda}l(l+1)\delta_{ll'}\delta_{mm'}, \\
& \oint T_{ab}^{B1,lm}(T^{ab}_{B1,l'm'})^*d\Omega=N^{(ten)}(B1,r,l)\delta_{ll'}\delta_{mm'}=\frac{1}{2}e^{-4\Lambda-2\Phi}l(l+1)\delta_{ll'}\delta_{mm'}, \\
& \oint T_{ab}^{E2,lm}(T^{ab}_{E2,l'm'})^*d\Omega=N^{(ten)}(E2,r,l)\delta_{ll'}\delta_{mm'}=\frac{(l+2)!}{2(l-2)!}\delta_{ll'}\delta_{mm'}, \\
& \oint T_{ab}^{B2,lm}(T^{ab}_{B2,l'm'})^*d\Omega=N^{(ten)}(B2,r,l)\delta_{ll'}\delta_{mm'}=e^{-2\Lambda-2\Phi}\frac{(l+2)!}{2(l-2)!}\delta_{ll'}\delta_{mm'}.
\end{align*}

Projecting the perturbed metric $h_{ab}$ onto these basis, one can obtain the expressions of coefficients A-K
\begin{align*}
& \mathrm{A}=e^{4\Phi}\oint h^{lm}_{ab}(v^av^bY^{\ast}_{lm})d\Omega,  &\mathrm{B}&=-\frac{e^{2\Phi}}{l(l+1)}\oint h^{lm}_{ab}v^{a}Y^{b\ast}_{E,lm}d\Omega, \\
& \mathrm{C}=-\frac{e^{4\Phi+2\Lambda}}{l(l+1)}\oint h^{lm}_{ab}v^{a}Y^{b\ast}_{B,lm}d\Omega,
& \mathrm{D}&=-e^{2\Phi+2\Lambda}\oint h^{lm}_{ab}v^{a}Y^{b\ast}_{R,lm}d\Omega,  \\
& \mathrm{E}=\frac{1}{2}\oint h_{ab}^{lm}T^{ab\ast}_{T0,lm}d\Omega, &\mathrm{F}&=\frac{2(l-2)!}{(l+2)!}\oint h_{ab}^{lm}T^{ab\ast}_{E2,lm}d\Omega, \\
& \mathrm{G}=\frac{2(l-2)!}{(l+2)!}e^{2\Phi+2\Lambda}\oint h_{ab}^{lm}T^{ab\ast}_{B2,lm}d\Omega,
& \mathrm{H}&=\frac{e^{2\Lambda}}{l(l+1)}\oint h^{lm}_{ab}T^{ab\ast}_{E1,lm}d\Omega, \\
& \mathrm{J}=\frac{e^{4\Lambda+2\Phi}}{l(l+1)}\oint h^{lm}_{ab}T^{ab\ast}_{B1,lm}d\Omega, &\mathrm{K}&=e^{4\Lambda}\oint h^{lm}_{ab}T^{ab\ast}_{L0,lm}d\Omega.
\end{align*}
The above results are only valid for $l\geq 2$.

\section{Decomposition of Linearized Einstein Equations}
In this appendix, we show the decomposition of the linearized Einstein tensor into the A-K components. Note that these expressions are only valid for $l\geq 2$.

\subsection{Even Parity}
\begin{align}
\label{projEA}
E_{\mathrm{A}} =& 2r^{-2}e^{-2\Lambda}(-1+e^{2\Lambda}+2r\Lambda')\mathrm{A}
-r^{-2}e^{2\Phi}(l^2+l-2)\mathrm{E} \non
& -2r^{-1}e^{-2\Lambda+2\Phi}(r\Lambda'-3)\frac{\pp }{\pp r}\mathrm{E}+2e^{-2\Lambda+2\Phi}\frac{\pp^2}{\pp r^2}\mathrm{E}
-\frac{1}{2}r^{-2}e^{2\Phi}l(l+2)(l+1)(l-1)\mathrm{F} \non
& -2r^{-2}e^{-2\Lambda+2\Phi}l(l+1)(r\Lambda'-2)\mathrm{H}+2r^{-1}e^{-2\Lambda+2\Phi}l(l+1)\frac{\pp }{\pp r}\mathrm{H} \non
& -r^{-2}e^{-4\Lambda+2\Phi}\left[2+e^{2\Lambda}l(l+1)-8r\Lambda' \right]\mathrm{K}-2r^{-1}e^{-4\Lambda+2\Phi}\frac{\pp }{\pp r}\mathrm{K} ,
\end{align}

\begin{align}
\label{projEB}
E_{\mathrm{B}} =& r^{-2}e^{-2\Lambda}(-2+2e^{2\Lambda}+3r\Lambda'-r\Phi'+2r^2\Lambda'\Phi'-2r^2{\Phi'}^2-2r^2\Phi'')\mathrm{B} \non
& -r^{-1}e^{-2\Lambda}(-2+r\Lambda'+r\Phi')\frac{\pp }{\pp r}\mathrm{B}+e^{-2\Lambda}\frac{\pp^2}{\pp r^2}\mathrm{B}
+r^{-1}e^{-2\Lambda}(\Lambda'-\Phi')\mathrm{D}-r^{-1}e^{-2\Lambda}\frac{\pp }{\pp r}\mathrm{D} \non
& -r^{-1}\frac{\pp }{\pp t}\mathrm{E}-\frac{1}{2}r^{-1}(l^2+l-2)\frac{\pp }{\pp t}\mathrm{F}-r^{-1}e^{-2\Lambda}(-3+r\Lambda'+r\Phi')\frac{\pp }{\pp t}\mathrm{H}
+e^{-2\Lambda}\frac{\pp^2}{\pp t\pp r}\mathrm{H} \non
& -r^{-1}e^{-2\Lambda}\frac{\pp }{\pp t}\mathrm{K} ,
\end{align}

\begin{align}
\label{projED}
E_{\mathrm{D}} =& r^{-2}l(l+1)(1-2r\Phi')\mathrm{B}+r^{-1}l(l+1)\frac{\pp}{\pp r}\mathrm{B}+2r^{-1}(-1+r\Phi')\frac{\pp}{\pp t}\mathrm{E}
-2\frac{\pp }{\pp t\pp r}\mathrm{E} \non
& -r^{-2}e^{-2\Lambda}(2-2e^{2\Lambda}+e^{2\Lambda}l+e^{2\Lambda}l^2-4r\Lambda')\mathrm{D}
-r^{-1}l(l+1)\frac{\pp }{\pp t}\mathrm{H}+2r^{-1}e^{-2\Lambda}\frac{\pp }{\pp t}\mathrm{K} ,
\end{align}

\begin{align}
\label{projEE}
E_{\mathrm{E}} =& -\frac{1}{2}r^{-2}e^{-2\Lambda-2\Phi}
(e^{2\Lambda}l+e^{2\Lambda}l^2+4r\Phi'-4r^2\Lambda'\Phi'+4r^2\Phi'')\mathrm{A}-r^{-1}e^{-2\Lambda-2\Phi}(-1+r\Lambda'+2r\Phi')\frac{\pp }{\pp r}\mathrm{A} \non
& +e^{-2\Lambda-2\Phi}\frac{\pp^2 }{\pp r^2}\mathrm{A}-r^{-1}e^{-2\Phi}l(l+1)\frac{\pp}{\pp t}\mathrm{B}
-2r^{-1}e^{-2\Lambda-2\Phi}(-1+r\Lambda')\frac{\pp}{\pp t}\mathrm{D}
+2e^{-2\Lambda-2\Phi}\frac{\pp^2}{\pp t\pp r}\mathrm{D} \non
& -2r^{-1}e^{-2\Lambda}(\Phi'-\Lambda'-r\Lambda'\Phi'+r{\Phi'}^2+r^2\Phi'')\mathrm{E}
+r^{-1}e^{-2\Lambda}(-2+r\Lambda'-r\Phi')\frac{\pp }{\pp r}\mathrm{E} \non
& -e^{-2\Lambda}\frac{\pp^2}{\pp r^2}\mathrm{E}+e^{-2\Phi}\frac{\pp^2 }{\pp t^2}\mathrm{E}
+r^{-2}e^{-2\Lambda}l(l+1)(-1+r\Lambda'-r\Phi')\mathrm{H}-r^{-1}e^{-2\Lambda}l(l+1)\frac{\pp }{\pp r}\mathrm{H} \non
& +\frac{1}{2}r^{-2}e^{-4\Lambda}(e^{2\Lambda}l+e^{2\Lambda}l^2+4r\Phi'-8r\Lambda'+4r^2{\Phi'}^2-8r^2\Lambda'\Phi'+4r^2\Phi'')\mathrm{K}
+r^{-1}e^{-4\Lambda}(1+r\Phi')\frac{\pp}{\pp r}\mathrm{K} \non
& +e^{-2\Lambda-2\Phi}\frac{\pp^2}{\pp t^2}\mathrm{K} ,
\end{align}

\begin{align}
\label{projEF}
E_{\mathrm{F}} =& -r^{-2}e^{-2\Phi}\mathrm{A}-2r^{-1}e^{-2\Phi}\frac{\pp }{\pp t}\mathrm{B}
-2r^{-1}e^{-2\Lambda}(\Phi'+r{\Phi'}^2-\Lambda'(1+r\Phi')+r\Phi'')\mathrm{F} \non
& -r^{-1}e^{-2\Lambda}(-2+r\Lambda'-r\Phi')\frac{\pp }{\pp r}\mathrm{F}+e^{-2\Lambda}\frac{\pp^2 }{\pp r^2}\mathrm{F}
-e^{-2\Phi}\frac{\pp^2 }{\pp t^2}\mathrm{F}+2r^{-2}e^{-2\Lambda}(-1+r\Lambda'-r\Phi')\mathrm{H} \non
& -2r^{-1}e^{-2\Lambda}\frac{\pp }{\pp r}\mathrm{H}+r^{-2}e^{-2\Lambda}\mathrm{K} ,
\end{align}

\begin{align}
\label{projEH}
E_{\mathrm{H}} =& r^{-2}e^{-2\Phi}(1+r\Phi')\mathrm{A}-r^{-1}e^{-2\Phi}\frac{\pp }{\pp r}\mathrm{A}+r^{-1}e^{-2\Phi}\frac{\pp }{\pp t}\mathrm{B}
-e^{-2\Phi}\frac{\pp^2 }{\pp t\pp r}\mathrm{B}-r^{-1}e^{-2\Phi}\frac{\pp}{\pp t}\mathrm{D} + r^{-1}\frac{\pp }{\pp r}\mathrm{E} \non
& +\frac{1}{2}r^{-1}(l+2)(l-1)\frac{\pp }{\pp r}\mathrm{F}+2r^{-2}e^{-2\Lambda}(e^{2\Lambda}-r\Phi'-r^2{\Phi'}^2+r\Lambda'(1+r\Phi')-r^2\Phi'')\mathrm{H} \non
& -e^{-2\Phi}\frac{\pp^2 }{\pp t^2}\mathrm{H} -r^{-2}e^{-2\Lambda}(1+r\Phi')\mathrm{K} ,
\end{align}

\begin{align}
\label{projEK}
E_{\mathrm{K}} =& -r^{-2}e^{-2\Phi}(e^{2\Lambda}l+e^{2\Lambda}l^2+4r\Phi')\mathrm{A}+2r^{-1}e^{-2\Phi}\frac{\pp }{\pp r}\mathrm{A}
-2r^{-1}e^{2\Lambda-2\Phi}l(l+1)\frac{\pp }{\pp t}\mathrm{B}+4r^{-1}e^{-2\Phi}\frac{\pp }{\pp t}\mathrm{D} \non
& +r^{-2}e^{2\Lambda}(l^2+l-2)\mathrm{E}-2r^{-1}(1+r\Phi')\frac{\pp }{\pp r}\mathrm{E}
+2e^{2\Lambda-2\Phi}\frac{\pp^2 }{\pp t^2}\mathrm{E}+\frac{1}{2}r^{-2}e^{2\Lambda}l(l+2)(l+1)(l-1)\mathrm{F} \non
& -2r^{-2}l(l+1)(1+r\Phi')\mathrm{H}+2r^{-2}\mathrm{K} .
\end{align}

\subsection{Odd Parity}

\begin{align}
\label{projEC}
E_{\mathrm{C}} =& -r^{-2}e^{-2\Lambda}\left(2-2e^{2\Lambda}+e^{2\Lambda}l+e^{2\Lambda}l^2-2r^2{\Lambda'}^2
+3r\Phi'-r\Lambda'(1+6r\Phi')+r^2\Lambda''+3r^2\Phi'' \right) \mathrm{C} \non
& -r^{-1}e^{-2\Lambda}(-2+3r\Lambda'+3r\Phi')\frac{\pp }{\pp r}\mathrm{C}+e^{-2\Lambda}\frac{\pp^2 }{\pp r^2}\mathrm{C}
-\frac{1}{2} r^{-1}(l+2)(l-1)\frac{\pp }{\pp t}\mathrm{G} \non
& -r^{-1}e^{-2\Lambda}(-3+2r\Lambda'+2r\Phi')\frac{\pp }{\pp t}\mathrm{J}+e^{-2\Lambda}\frac{\pp^2 }{\pp t\pp r}\mathrm{J} ,
\end{align}

\begin{align}
\label{projEG}
E_{\mathrm{G}} &= -2r^{-1}e^{-2\Phi}\frac{\pp }{\pp t}\mathrm{C}
+r^{-1}e^{-2\Lambda}(2r{\Lambda'}^2-4\Phi'+4r\Lambda'\Phi'-2r{\Phi'}^2-r\Lambda''-3r\Phi'')\mathrm{G} \non
& -r^{-1}e^{-2\Lambda}(-2+3r\Lambda'+r\Phi')\frac{\pp }{\pp r}\mathrm{G}+e^{-2\Lambda}\frac{\pp^2 }{\pp r^2}\mathrm{G}-e^{-2\Phi}\frac{\pp^2 }{\pp t^2}\mathrm{G}
+2r^{-2}e^{-2\Lambda}(-1+2r\Lambda')\mathrm{J}-2r^{-1}e^{-2\Lambda}\frac{\pp }{\pp r}\mathrm{J} ,
\end{align}

\begin{align}
\label{projEJ}
E_{\mathrm{J}} &= r^{-1}e^{-2\Phi}(1+r\Lambda'+r\Phi')\frac{\pp }{\pp t}\mathrm{C}-e^{-2\Phi}\frac{\pp^2 }{\pp t\pp r}\mathrm{C}
-\frac{1}{2}r^{-1}(l+2)(l-1)(\Lambda'+\Phi')\mathrm{G}+\frac{1}{2}r^{-1}(l+2)(l-1)\frac{\pp }{\pp r}\mathrm{G}  \non
& -r^{-2}e^{-2\Lambda}(-2e^{2\Lambda}+e^{2\Lambda}l+e^{2\Lambda}l^2-2r\Lambda'+2r\Phi'+2r^2{\Phi'}^2+2r^2\Phi''-2r^2\Lambda'\Phi')\mathrm{J}
-e^{-2\Phi}\frac{\pp^2 }{\pp t^2}\mathrm{J} .
\end{align}

\section{Linearized Einstein Tensor In Terms of Gauge Invariants}
In this appendix, we rewrite the linearized Einstein tensor in terms of the gauge invariants under the EZ gauge or the RW gauge. Note that for even-parity perturbations, we use the superscripts to distinguish expressions under the EZ and RW gauges, respectively. These relations are only valid for $l\geq 2$ case. The relation $\lambda=l(l+1)$ also has been used in this appendix.

\subsection{Even Parity and the EZ Gauge}

\begin{align}
\label{projEAgaugeinEZ}
E_\mathrm{A}^{\mathrm{I}} =& -4r^{-2}e^{-2\Lambda}(-1+e^{2\Lambda}+2r\Lambda')\epsilon^\mathrm{I}
-2r^{-2}e^{-2\Lambda+2\Phi}\lambda(r\Lambda'-2)\chi^\mathrm{I}+2r^{-1}e^{-2\Lambda+2\Phi}\lambda\frac{\pp }{\pp r}\chi^\mathrm{I} \non
& -2r^{-2}e^{-4\Lambda+2\Phi}\left(2+e^{2\Lambda}\lambda-8r\Lambda'\right)\psi^\mathrm{I}
-4r^{-1}e^{-4\Lambda+2\Phi}\frac{\pp }{\pp r}\psi^\mathrm{I},
\end{align}

\begin{align}
\label{projEBgaugeinEZ}
E_\mathrm{B}^{\mathrm{I}} =& r^{-1}e^{-2\Lambda}(\Lambda'-\Phi')\delta^\mathrm{I}-r^{-1}e^{-2\Lambda}\frac{\pp }{\pp r}\delta^\mathrm{I}
-r^{-1}e^{-2\Lambda}(-3+r\Lambda'+r\Phi')\frac{\pp}{\pp t}\chi^\mathrm{I}
+e^{-2\Lambda}\frac{\pp^2}{\pp t\pp r}\chi^\mathrm{I}-2r^{-1}e^{-2\Lambda}\frac{\pp }{\pp t}\psi^\mathrm{I} ,
\end{align}

\begin{align}
\label{projEDgaugeinEZ}
E_\mathrm{D}^{\mathrm{I}} =& -r^{-2}e^{-2\Lambda}\left[2+e^{2\Lambda}(\lambda-2)-4r\Lambda'\right]\delta^\mathrm{I}
-r^{-1}\lambda\frac{\pp }{\pp t}\chi^\mathrm{I}+4r^{-1}e^{-2\Lambda}\frac{\pp }{\pp t}\psi^\mathrm{I} ,
\end{align}

\begin{align}
\label{projEEgaugeinEZ}
E_\mathrm{E}^{\mathrm{I}} =& r^{-2}e^{-2\Lambda-2\Phi}\left(e^{2\Lambda}\lambda+4r\Phi'-4r^2\Lambda'\Phi'+4r^2\Phi''\right)\epsilon^\mathrm{I}
+2r^{-1}e^{-2\Lambda-2\Phi}(-1+r\Lambda'+2r\Phi')\frac{\pp }{\pp r}\epsilon^\mathrm{I} \non
& -2e^{-2\Lambda-2\Phi}\frac{\pp^2 }{\pp r^2}\epsilon^\mathrm{I}-2r^{-1}e^{-2\Lambda-2\Phi}(-1+r\Lambda')\frac{\pp}{\pp t}\delta^\mathrm{I}
+2e^{-2\Lambda-2\Phi}\frac{\pp^2}{\pp t\pp r}\delta^\mathrm{I}+r^{-2}e^{-2\Lambda}\lambda(-1+r\Lambda'-r\Phi')\chi^\mathrm{I} \non
& -r^{-1}e^{-2\Lambda}\lambda\frac{\pp }{\pp r}\chi^\mathrm{I}
+r^{-2}e^{-4\Lambda}\left(e^{2\Lambda}\lambda+4r\Phi'-8r\Lambda'+4r^2{\Phi'}^2-8r^2\Lambda'\Phi'+4r^2\Phi''\right)\psi^\mathrm{I} \non
& +2r^{-1}e^{-4\Lambda}(1+r\Phi')\frac{\pp }{\pp r}\psi^\mathrm{I} +2e^{-2\Lambda-2\Phi}\frac{\pp^2 }{\pp t^2}\psi^\mathrm{I} ,
\end{align}

\begin{align}
\label{projEFgaugeinEZ}
E_\mathrm{F}^\mathrm{I} =& 2r^{-2}e^{-2\Phi}\epsilon^\mathrm{I}+2r^{-2}e^{-2\Lambda}(-1+r\Lambda'-r\Phi')\chi^\mathrm{I}
-2r^{-1}e^{-2\Lambda}\frac{\pp }{\pp r}\chi^\mathrm{I}+2r^{-2}e^{-2\Lambda}\psi^\mathrm{I} ,
\end{align}

\begin{align}
\label{projEHgaugeinEZ}
E_\mathrm{H}^\mathrm{I} =& -2r^{-2}e^{-2\Phi}(1+r\Phi')\epsilon^\mathrm{I}+2r^{-1}e^{-2\Phi}\frac{\pp }{\pp r}\epsilon^\mathrm{I}
-r^{-1}e^{-2\Phi}\frac{\pp }{\pp t}\delta^\mathrm{I}-2r^{-2}e^{-2\Lambda}(1+r\Phi')\psi^\mathrm{I} \non
& +2r^{-2}e^{-2\Lambda}(e^{2\Lambda}-r\Phi'-r^2{\Phi'}^2+r\Lambda'(1+r\Phi')-r^2\Phi'')\chi^\mathrm{I}
-e^{-2\Phi}\frac{\pp^2 }{\pp t^2}\chi^\mathrm{I} ,
\end{align}

\begin{align}
\label{projEKgaugeinEZ}
E_\mathrm{K}^\mathrm{I} =& 2r^{-2}e^{-2\Phi}\left(e^{2\Lambda}\lambda+4r\Phi'\right)\epsilon^\mathrm{I}-4r^{-1}e^{-2\Phi}\frac{\pp }{\pp r}\epsilon^\mathrm{I}
+4r^{-1}e^{-2\Phi}\frac{\pp}{\pp t}\delta^\mathrm{I}-2r^{-2}\lambda(1+r\Phi')\chi^\mathrm{I}+4r^{-2}\psi^\mathrm{I} .
\end{align}

\subsection{Even Parity and the RW Gauge}

\begin{align}
\label{projEAgaugeinRW}
E_\mathrm{A}^\mathrm{II} =&
-4r^{-2}e^{-2\Lambda}\left(-1+e^{2\Lambda}+2r\Lambda'\right)\epsilon^\mathrm{II} \non
& +r^{-2}e^{-4\Lambda+2\Phi} \left[ 4-4e^{2\Lambda}+4e^{2\Lambda}\lambda-8r^2{\Lambda'}^2
-2r \Lambda'\left(e^{2\Lambda}\lambda-4r\Phi'\right)+4r^2\Lambda''-4r\Phi'(1-e^{2\Lambda}) \right]\chi^\mathrm{II} \non
& +2r^{-1}e^{-2\Lambda+2\Phi}\lambda\frac{\pp }{\pp r}\chi-2r^{-2}e^{-4\Lambda+2\Phi}(2+e^{2\Lambda}\lambda-8r\Lambda')\psi^\mathrm{II}
-4r^{-1}e^{-4\Lambda+2\Phi}\frac{\pp}{\pp r}\psi^\mathrm{II} ,
\end{align}

\begin{align}
\label{projEBgaugeinRW}
E_\mathrm{B}^\mathrm{II} = r^{-1}e^{-2\Lambda}(\Lambda'-\Phi')\delta^\mathrm{II}-r^{-1}e^{-2\Lambda}\frac{\pp }{\pp r}\delta^\mathrm{II}
-r^{-1}e^{-2\Lambda}(-3+r\Lambda'+r\Phi')\frac{\pp }{\pp t}\chi^\mathrm{II}
+e^{-2\Lambda}\frac{\pp^2}{\pp t\pp r}\chi^\mathrm{II}-2r^{-1}e^{-2\Lambda}\frac{\pp }{\pp t}\psi^\mathrm{II} , \non
\end{align}

\begin{align}
\label{projEDgaugeinRW}
E_\mathrm{D}^\mathrm{II} =& -r^{-2}e^{-2\Lambda}\left(2-2e^{2\Lambda}+e^{2\Lambda}\lambda-4r\Lambda'\right)\delta^\mathrm{II}
-r^{-1}e^{-2\Lambda}\left(2-2e^{2\Lambda}+e^{2\Lambda}\lambda+4r\Phi'\right)\frac{\pp}{\pp t}\chi^\mathrm{II} +4r^{-1}e^{-2\Lambda}\frac{\pp }{\pp t}\psi^\mathrm{II} ,
\end{align}

\begin{align}
\label{projEEgaugeinRW}
E_\mathrm{E}^\mathrm{II} =& r^{-2}e^{-2\Lambda-2\Phi} \left(e^{2\Lambda}\lambda+4r\Phi'-4r^2\Lambda'\Phi'+4r^2\Phi''\right)\epsilon^\mathrm{II}
+2r^{-1}e^{-2\Lambda-2\Phi}(-1+r\Lambda'+2r\Phi')\frac{\pp }{\pp r}\epsilon^\mathrm{II} \non
& -2e^{-2\Lambda-2\Phi}\frac{\pp^2 }{\pp r^2}\epsilon^\mathrm{II}- 2r^{-1}e^{-2\Lambda-2\Phi}(-1+r\Lambda')\frac{\pp  }{\pp t}\delta^\mathrm{II}
+2e^{-2\Lambda-2\Phi}\frac{\pp^2 }{\pp t\pp r}\delta^\mathrm{II} \non
& +r^{-2}e^{-4\Lambda}\left[-e^{2\Lambda}\lambda+4r^2{\Phi'}^2+4r^2{\Lambda'}^2(1+r\Phi')-2r^2\Lambda''+6r^2\Phi'' \right. \non
& \left. -r\Phi'(-2+e^{2\Lambda}\lambda+2r^2\Lambda''-4r^2\Phi'')
-r\Lambda'(2-e^{2\Lambda}\lambda+8r\Phi'+4r^2{\Phi'}^2+6r^2\Phi'')+2r^3\Phi'''\right]\chi^\mathrm{II} \non
& -r^{-1}e^{-2\Lambda}\lambda\frac{\pp }{\pp r}\chi^\mathrm{II}
+r^{-2}e^{-4\Lambda}\left(e^{2\Lambda}\lambda+4r\Phi'-8r\Lambda'+4r^2{\Phi'}^2-8r^2\Lambda'\Phi'+4r^2\Phi''\right)\psi^\mathrm{II} \non
& +2r^{-1}e^{-4\Lambda}(1+r\Phi')\frac{\pp }{\pp r}\psi^\mathrm{II}+2e^{-2\Lambda-2\Phi}\frac{\pp^2 }{\pp t^2}\psi^\mathrm{II} ,
\end{align}

\begin{align}
\label{projEFgaugeinRW}
E_\mathrm{F}^\mathrm{II} =& 2r^{-2}e^{-2\Phi}\epsilon^\mathrm{II}+2r^{-2}e^{-2\Lambda}(-1+r\Lambda'-r\Phi')\chi^\mathrm{II}
-2r^{-1}e^{-2\Lambda}\frac{\pp }{\pp r}\chi^\mathrm{II}+2r^{-2}e^{-2\Lambda}\psi^\mathrm{II} ,
\end{align}

\begin{align}
\label{projEHgaugeinRW}
E_\mathrm{H}^\mathrm{II} =& -2r^{-2}e^{-2\Phi}(1+r\Phi')\epsilon^\mathrm{II}+2r^{-1}e^{-2\Phi}\frac{\pp }{\pp r}\epsilon^\mathrm{II}
-r^{-1}e^{-2\Phi}\frac{\pp }{\pp t}\delta^\mathrm{II}-2r^{-2}e^{-2\Lambda}(1+r\Phi')\psi^\mathrm{II} \non
& +2r^{-2}e^{-2\Lambda}\left(1+r\Phi'+r\Lambda'-r^2{\Phi'}^2+r^2\Lambda'\Phi'-r^2\Phi''\right)\chi^\mathrm{II}
-e^{-2\Phi}\frac{\pp^2 }{\pp t^2}\chi^\mathrm{II} ,
\end{align}

\begin{align}
\label{projEKgaugeinRW}
E_\mathrm{K}^\mathrm{II} =& 2r^{-2}e^{-2\Phi}(e^{2\Lambda}\lambda+4r\Phi')\epsilon^\mathrm{II}-4r^{-1}e^{-2\Phi}\frac{\pp }{\pp r}\epsilon^\mathrm{II}
+4r^{-1}e^{-2\Phi}\frac{\pp }{\pp t}\delta^\mathrm{II} \non
& -2r^{-2}e^{-2\Lambda}\left[e^{2\Lambda}\lambda+r\Phi'(e^{2\Lambda}\lambda-2)+2r\Lambda'(2-e^{2\Lambda}+4r\Phi')-2r^2\Phi'' \right]\chi^\mathrm{II} \non
& +4r^{-1}e^{-2\Lambda}\left(1-e^{2\Lambda}+2r\Phi'\right)\frac{\pp}{\pp r}\chi^\mathrm{II}
+4r^{-2}e^{-2\Lambda}\left(e^{2\Lambda}+2r\Lambda'+r^2\Lambda'\Phi'-r^2{\Phi'}^2-r^2\Phi''\right)\psi^\mathrm{II} .
\end{align}

\subsection{Odd Parity}

\begin{align}
\label{projECgaugein}
E_\mathrm{C} =& r^{-2}e^{-2\Lambda} \left[2-2e^{2\Lambda}+e^{2\Lambda}\lambda-2r^2{\Lambda'}^2
+3r\Phi'-r\Lambda'(1+6r\Phi')+r^2\Lambda''+3r^2\Phi'' \right]\beta \non
& +r^{-1}e^{-2\Lambda}(-2+3r\Lambda'+3r\Phi')\frac{\pp}{\pp r}\beta-e^{-2\Lambda}\frac{\pp^2}{\pp r^2}\beta
-r^{-1}e^{-2\Lambda}(-3+2r\Lambda'+2r\Phi')\frac{\pp}{\pp t}\alpha+e^{-2\Lambda}\frac{\pp^2}{\pp t\pp r}\alpha ,
\end{align}

\begin{align}
\label{projEGgaugein}
E_\mathrm{G} =& 2r^{-1}e^{-2\Phi}\frac{\pp }{\pp t}\beta+2r^{-2}e^{-2\Lambda}(-1+2r\Lambda')\alpha-2r^{-1}e^{-2\Lambda}\frac{\pp }{\pp r}\alpha ,
\end{align}

\begin{align}
\label{projEJgaugein}
E_\mathrm{J} =& -r^{-2}e^{-2\Lambda}\left[e^{2\Lambda}\lambda-2e^{2\Lambda}+2r\Phi'+2r^2{\Phi'}^2
+2r^2\Phi''-2r\Lambda'(1+r\Phi')\right]\alpha \non
& -e^{-2\Phi}\frac{\pp^2 }{\pp t^2}\alpha-r^{-1}e^{-2\Phi}(1+r\Lambda'+r\Phi')\frac{\pp }{\pp t}\beta+e^{-2\Phi}\frac{\pp^2 }{\pp t\pp r}\beta .
\end{align}

\section{The specific structure of the parameters}

\subsection{Even-parity and the EZ gauge for \texorpdfstring{$l\geq 2$}.}

The functions $\sigma^\mathrm{I}$, $\tau^\mathrm{I}$, $\eta^\mathrm{I}$, $\gamma^\mathrm{I}$, $\rho^\mathrm{I}$, $\mu^\mathrm{I}$, $\nu^\mathrm{I}$, $\kappa^\mathrm{I}$ in \eqs{prchiEZ} and \meq{prpsiEZ} have the following form

\begin{align}
\sigma^\mathrm{I} =& 1+e^{2\Lambda}(\lambda-1)-2r\Lambda' ,\\
\tau^\mathrm{I} =& -2+e^{2 \Lambda}\lambda +2r\Phi' ,\\
\eta^\mathrm{I} =& 2+e^{2\Lambda}(\lambda-2)-4r\Lambda' ,\\
\gamma^\mathrm{I} =& 2-2e^{2\Lambda}+r\left(e^{2 \Lambda}\lambda-4\right)\Lambda'+2r\Phi'+2r^2\Phi'' ,\\
\rho^\mathrm{I} =& e^{2\Lambda} (\lambda-2)+4r\Phi' ,\\
\mu^{\mathrm{I}} =& -2 e^{2\Lambda}+2e^{4\Lambda}-2e^{4\Lambda}\lambda+e^{4\Lambda}\lambda^2-r\left(-2+e^{2\Lambda}(2-4\lambda)+e^{4 \Lambda }\lambda\right)\Phi'
-2\left(-1+e^{2 \Lambda}\right)r^2{\Phi'}^2+4 r^2{\Lambda'}^2\left(1+r\Phi'\right) \non
& +2r^2\Phi''-2e^{2\Lambda}r^2\Phi''+e^{2\Lambda}r^2\lambda\Phi''-r\Lambda'\left(2-6e^{2\Lambda}+3e^{2\Lambda}\lambda
+r\left(6+e^{2\Lambda}(-2+3\lambda )\right)\Phi'+4r^2{\Phi'}^2+4r^2\Phi''\right) ,\\
\nu^\mathrm{I} =& 4-4e^{2\Lambda}-e^{4\Lambda}(\lambda-2)+e^{2\Lambda}r(\lambda-2)\Phi'+4r\Lambda'\left(-3+e^{2\Lambda}(1+\lambda)+r\Phi' \right),\\
\kappa^{\mathrm{I}} =& -e^{2\Lambda}\lambda \left(e^{2\Lambda}\lambda-2+2r\Phi' \right).
\end{align}
To simplify the expression of $N_\mathrm{R}$ and $N_\mathrm{Z}$, we first introduce two auxiliary functions $M^\mathrm{I}$ and $M^\mathrm{II}$
\begin{align}
M^{\mathrm{I}}_1 =& -2 {\nu^{\mathrm{I}}}+{\rho^{\mathrm{I}}} \sigma^{\mathrm{I}},\\
M^{\mathrm{I}}_2 =& \frac{2 }{\sigma^{\mathrm{I}}}\left(-2 {\mu^{\mathrm{I}}}+{\gamma^{\mathrm{I}}} \sigma^{\mathrm{I}}+r {\tau^{\mathrm{I}}} \left({\sigma^{\mathrm{I}}}\right)'\right).
\end{align}
Then the functions in \eq{evenmasterEZ} are given by
\begin{equation}
{N^{\mathrm{I}}} =-4 {\mu^{\mathrm{I}}}+{\sigma^{\mathrm{I}}} (2 {\gamma^{\mathrm{I}}}-2 {\nu^{\mathrm{I}}}+{\rho^{\mathrm{I}}} {\sigma^{\mathrm{I}}})+2 r {\tau^{\mathrm{I}}}
\left({\sigma^{\mathrm{I}}}\right)',
\end{equation}
\begin{align}
N^{\mathrm{I}}_\mathrm{R} =&-\left({N^{\mathrm{I}}}\right)^2-2 r {\sigma^{\mathrm{I}}} {\tau^{\mathrm{I}}} \left({N^{\mathrm{I}}}\right)'+{N^{\mathrm{I}}} \left(-4 {\mu^{\mathrm{I}}}+{M^{\mathrm{I}}_1} {\sigma^{\mathrm{I}}}-2
{\nu^{\mathrm{I}}} {\sigma^{\mathrm{I}}}+2 {\sigma^{\mathrm{I}}} {\tau^{\mathrm{I}}}+2 r {\tau^{\mathrm{I}}} \left({\sigma^{\mathrm{I}}}\right)'+2 r {\sigma^{\mathrm{I}}} \left({\tau^{\mathrm{I}}}\right)'\right),\\
N^{\mathrm{I}}_\mathrm{T} =&-\left({N^{\mathrm{I}}}\right)^2 \left({\eta^{\mathrm{I}}}\right)^2+2
{N^{\mathrm{I}}} \left(r \left({\sigma^{\mathrm{I}}}\right)^2 {\tau^{\mathrm{I}}} \left({\eta^{\mathrm{I}}}\right)'+{\eta^{\mathrm{I}}} {\sigma^{\mathrm{I}}} \left(2 {\mu^{\mathrm{I}}}+{\nu^{\mathrm{I}}} \sigma^{\mathrm{I}}-2 {\tau^{\mathrm{I}}} \left(r \left({\sigma^{\mathrm{I}}}\right)'+{\sigma^{\mathrm{I}}} \left(1+r \Lambda '-r \Phi '\right)\right)\right)\right.\non
&\left.+2 r {\eta^{\mathrm{I}}} {\sigma^{\mathrm{I}}} (-{\eta^{\mathrm{I}}}+{\sigma^{\mathrm{I}}}) {\tau^{\mathrm{I}}} \left({N^{\mathrm{I}}}\right)'+\left({\eta^{\mathrm{I}}}\right)^2 \left(-2 \mu^{\mathrm{I}}-{\nu^{\mathrm{I}}} {\sigma^{\mathrm{I}}}+{\tau^{\mathrm{I}}} \left(r \left({\sigma^{\mathrm{I}}}\right)'+2 {\sigma^{\mathrm{I}}} \left(1+r \Lambda '-r \Phi '\right)\right)\right)\right),\\
N^{\mathrm{I}}_\mathrm{Z} =&{N^{\mathrm{I}}} {\sigma^{\mathrm{I}}} \left({M^{\mathrm{I}}_2} {\nu^{\mathrm{I}}}+r {\tau^{\mathrm{I}}} \left({M^{\mathrm{I}}_1}\right)'\right)+{M^{\mathrm{I}}_1} \left(-r \sigma^{\mathrm{I}} {\tau^{\mathrm{I}}} \left({N^{\mathrm{I}}}\right)'+{N^{\mathrm{I}}} \left(-2 {\mu^{\mathrm{I}}}+r {\tau^{\mathrm{I}}} \left({\sigma^{\mathrm{I}}}\right)'\right)\right),
\end{align}

and the functions in \eq{evensourceEZ} are given by
\begin{align}
N^{\mathrm{I}}_\mathrm{A} =&-\left({N^{\mathrm{I}}}\right)^2-2 r {\sigma^{\mathrm{I}}} {\tau^{\mathrm{I}}} \left({N^{\mathrm{I}}}\right)'+2 {N^{\mathrm{I}}} \left(-2 {\mu^{\mathrm{I}}}-{\nu^{\mathrm{I}}} \sigma^{\mathrm{I}}+2 {\sigma^{\mathrm{I}}} {\tau^{\mathrm{I}}}+4 r {\sigma^{\mathrm{I}}} {\tau^{\mathrm{I}}} \Lambda '+r {\tau^{\mathrm{I}}} \left({\sigma^{\mathrm{I}}}\right)'+r {\sigma^{\mathrm{I}}}
\left({\tau^{\mathrm{I}}}\right)'-2 r {\sigma^{\mathrm{I}}} {\tau^{\mathrm{I}}} \Phi '\right),\\
N^{\mathrm{I}}_\mathrm{D} =&\left({N^{\mathrm{I}}}\right)^2 \left({\eta^{\mathrm{I}}}\right)^2+2{N^{\mathrm{I}}} \left(-r \left({\sigma^{\mathrm{I}}}\right)^2 {\tau^{\mathrm{I}}} \left({\eta^{\mathrm{I}}}\right)'+\left({\eta^{\mathrm{I}}}\right)^2 \left(2 {\mu^{\mathrm{I}}}+{\nu^{\mathrm{I}}} {\sigma^{\mathrm{I}}}-\tau^{\mathrm{I}} \left(r \left({\sigma^{\mathrm{I}}}\right)'+{\sigma^{\mathrm{I}}} \left(3+4 r \Lambda '-2 r \Phi '\right)\right)\right)\right.\non
&\left.+2 r {\eta^{\mathrm{I}}} ({\eta^{\mathrm{I}}}-{\sigma^{\mathrm{I}}}) {\sigma^{\mathrm{I}}} {\tau^{\mathrm{I}}} \left({N^{\mathrm{I}}}\right)'
+{\eta^{\mathrm{I}}} {\sigma^{\mathrm{I}}} \left(-2
{\mu^{\mathrm{I}}}-{\nu^{\mathrm{I}}} {\sigma^{\mathrm{I}}}+{\tau^{\mathrm{I}}} \left(2 r \left({\sigma^{\mathrm{I}}}\right)'+{\sigma^{\mathrm{I}}} \left(3+4 r \Lambda '-2 r \Phi '\right)\right)\right)\right),\\
M^{\mathrm{I}}_\mathrm{F} =&\frac{1}{2} \left({\kappa^{\mathrm{I}}}+\sigma^{\mathrm{I}} \left(-2+e^{2 \Lambda } \lambda +2 r \Phi '\right)\right), \\
N^{\mathrm{I}}_\mathrm{F} =&\left({N^{\mathrm{I}}}\right)^2 {\kappa^{\mathrm{I}}}+4 r {M^{\mathrm{I}}_\mathrm{F}} {\sigma^{\mathrm{I}}} {\tau^{\mathrm{I}}} \left({N^{\mathrm{I}}}\right)'+4 {N^{\mathrm{I}}} \left(-r {\sigma^{\mathrm{I}}}
{\tau^{\mathrm{I}}} \left({M^{\mathrm{I}}_\mathrm{F}}\right)'+{M^{\mathrm{I}}_\mathrm{F}} \left(2 {\mu^{\mathrm{I}}}+{\nu^{\mathrm{I}}} {\sigma^{\mathrm{I}}}-{\tau^{\mathrm{I}}} \left(2 {\sigma^{\mathrm{I}}} \left(1+r
\Lambda '\right)+r \left({\sigma^{\mathrm{I}}}\right)'\right)\right)\right),\\
N^{\mathrm{I}}_{\mathrm{HK}}(r) =&\left({N^{\mathrm{I}}}\right)^2 {\eta^{\mathrm{I}}}+2 r ({\eta^{\mathrm{I}}}-{\sigma^{\mathrm{I}}}) {\sigma^{\mathrm{I}}} {\tau^{\mathrm{I}}} \left({N^{\mathrm{I}}}\right)'+2 {N^{\mathrm{I}}}
\left(-{\sigma^{\mathrm{I}}} \left(2 {\mu^{\mathrm{I}}}+{\nu^{\mathrm{I}}} {\sigma^{\mathrm{I}}}+{\tau^{\mathrm{I}}} \left(-2 {\sigma^{\mathrm{I}}} \left(1+r \Lambda '\right)+r
\left(\left({\eta^{\mathrm{I}}}\right)'-2 \left({\sigma^{\mathrm{I}}}\right)'\right)\right)\right)\right.\non
&\left.+{\eta^{\mathrm{I}}} \left(2 {\mu^{\mathrm{I}}}+{\nu^{\mathrm{I}}} {\sigma^{\mathrm{I}}}-\tau^{\mathrm{I}} \left(2 {\sigma^{\mathrm{I}}} \left(1+r \Lambda '\right)+r \left({\sigma^{\mathrm{I}}}\right)'\right)\right)\right).
\end{align}

\subsection{Even-parity and the RW gauge for \texorpdfstring{$l\geq 2$}.}

The functions $\sigma^\mathrm{II}$, $\tau^\mathrm{II}$, $\eta^\mathrm{II}$, $\gamma^\mathrm{II}$, $\rho^\mathrm{II}$, $\mu^\mathrm{II}$, $\nu^\mathrm{II}$ and $\kappa^\mathrm{II}$ in \eqs{prchiRW} and \meq{prpsiRW} have the following form

\begin{align}
\sigma^{\mathrm{II}} =&2-2 e^{2 \Lambda }+e^{2 \Lambda } \lambda -2 r \Lambda '+2 r \Phi ',\\
\tau^{\mathrm{II}} =&e^{2 \Lambda } (-2+\lambda )+6 r \Phi ',\\
\eta^{\mathrm{II}} =&2+e^{2 \Lambda } (-2+\lambda )-4 r \Lambda ',\\
\gamma^{\mathrm{II}} =&r \left(-4 \Phi '+\Lambda ' \left(e^{2 \Lambda } (-2+\lambda )+8 r \Phi '\right)\right),\\
\rho^{\mathrm{II}} =&e^{2 \Lambda } (-2+\lambda )+4 r \Phi ',\\
\mu^{\mathrm{II}} =&-2 r^2 \Lambda '^2 \left(e^{2 \Lambda } (-2+\lambda )+8 r \Phi '\right)+r \Lambda ' \left(-2 e^{2 \Lambda } (-2+\lambda )+r
\left(-2+e^{2 \Lambda } (-2+\lambda )\right) \Phi '\right)\non
&+e^{2 \Lambda } (-2+\lambda ) \left(2+e^{2 \Lambda } (-2+\lambda )+r^2 \Lambda ''\right)+2
r \Phi ' \left(4+2 e^{2 \Lambda } (-2+\lambda )+3 r^2 \Lambda ''\right),\\
\nu^{\mathrm{II}} =&-e^{2 \Lambda } (-2+\lambda )-r \left(8+e^{2 \Lambda } (-2+\lambda )\right) \Phi '+4 r \Lambda ' \left(e^{2 \Lambda } (-2+\lambda
)+7 r \Phi '\right),\\
\kappa^{\mathrm{II}} =&4-8 e^{2 \Lambda }+4 e^{4 \Lambda }+2 e^{2 \Lambda } \lambda -e^{4 \Lambda } \lambda ^2-2 r \left(-4+e^{2 \Lambda } (4+\lambda
)\right) \Phi '-8 r \Lambda ' \left(1-e^{2 \Lambda }+2 r \Phi '\right).
\end{align}
To simplify the expressions of $N_T^\mathrm{II}$, $N_R^\mathrm{II}$ and $N_Z^\mathrm{II}$, two auxiliary functions $M_1^\mathrm{II}$ and $M_2^\mathrm{II}$ are introduced
\begin{align}
M^{\mathrm{II}}_1 =&2 {\nu^{\mathrm{II}}}-{\rho^{\mathrm{II}}} {\sigma^{\mathrm{II}}},\\
M^{\mathrm{II}}_2 =&2 \left(-2 {\mu^{\mathrm{II}}}+{\gamma^{\mathrm{II}}} {\sigma^{\mathrm{II}}}+r {\tau^{\mathrm{II}}} \left({\sigma^{\mathrm{II}}}\right)'\right) .
\end{align}
Then the functions in \eq{evenmasterRW} are given by
\begin{align}
N^{\mathrm{II}} =&4 {\mu^{\mathrm{II}}}-{\sigma^{\mathrm{II}}} (2 {\gamma^{\mathrm{II}}}-2 {\nu^{\mathrm{II}}}+{\rho^{\mathrm{II}}} {\sigma^{\mathrm{II}}})-2 r \tau^{\mathrm{II}} \left({\sigma^{\mathrm{II}}}\right)',\\
N^{\mathrm{II}}_\mathrm{R} =&-\left({N^{\mathrm{II}}}\right)^2+2 r {\sigma^{\mathrm{II}}} {\tau^{\mathrm{II}}} \left({N^{\mathrm{II}}}\right)'+{N^{\mathrm{II}}} \left(4 {\mu^{\mathrm{II}}}+{M^{\mathrm{II}}_1} \sigma^{\mathrm{II}}+2 {\nu^{\mathrm{II}}} {\sigma^{\mathrm{II}}}-2 r {\tau^{\mathrm{II}}} \left({\sigma^{\mathrm{II}}}\right)'-2 {\sigma^{\mathrm{II}}} \left({\tau^{\mathrm{II}}}+r \left({\tau^{\mathrm{II}}}\right)'\right)\right),\\
N^{\mathrm{II}}_\mathrm{T} =&-\left({N^{\mathrm{II}}}\right)^2 \left({\eta^{\mathrm{II}}}\right)^2-4 r^2 {\eta^{\mathrm{II}}} {\sigma^{\mathrm{II}}} {\tau^{\mathrm{II}}} \left({N^{\mathrm{II}}}\right)' \left(\Lambda '+\Phi
'\right)+4 r {N^{\mathrm{II}}} \left(-r {\sigma^{\mathrm{II}}} {\tau^{\mathrm{II}}} \left({\eta^{\mathrm{II}}}\right)' \left(\Lambda '+\Phi '\right)\right.\non
&\left.+{\eta^{\mathrm{II}}} \left(-2 \mu^{\mathrm{II}} \left(\Lambda '+\Phi '\right)-{\nu^{\mathrm{II}}} {\sigma^{\mathrm{II}}} \left(\Lambda '+\Phi '\right)+{\tau^{\mathrm{II}}} \left(r \left({\sigma^{\mathrm{II}}}\right)'
\left(\Lambda '+\Phi '\right)\right.\right.\right.\non
&\left.\left.\left.+{\sigma^{\mathrm{II}}} \left(3 \Lambda '+2 r \Lambda '^2+3 \Phi '-2 r \Phi '^2+r \left(\Lambda ''+\Phi ''\right)\right)\right)\right)\right),\\
N^{\mathrm{II}}_\mathrm{Z} =&{N^{\mathrm{II}}} \left(-{M^{\mathrm{II}}_2} {\nu^{\mathrm{II}}}+r {\sigma^{\mathrm{II}}} {\tau^{\mathrm{II}}} \left({M^{\mathrm{II}}_1}\right)'\right)+{M^{\mathrm{II}}_1} \left(-r
{\sigma^{\mathrm{II}}} {\tau^{\mathrm{II}}} \left({N^{\mathrm{II}}}\right)'+{N^{\mathrm{II}}} \left(-2 {\mu^{\mathrm{II}}}+r {\tau^{\mathrm{II}}} \left({\sigma^{\mathrm{II}}}\right)'\right)\right).
\end{align}
and the functions in \eq{evensourceRW} are given by
\begin{align}
N^{\mathrm{II}}_{\mathrm{A}} =&2 {N^{\mathrm{II}}} \left(2 {\mu^{\mathrm{II}}}+{\nu^{\mathrm{II}}} \sigma^{\mathrm{II}}-2 {\sigma^{\mathrm{II}}} {\tau^{\mathrm{II}}}-4 r {\sigma^{\mathrm{II}}} {\tau^{\mathrm{II}}} \Lambda '-r {\tau^{\mathrm{II}}} \left({\sigma^{\mathrm{II}}}\right)'-r \sigma^{\mathrm{II}} \left({\tau^{\mathrm{II}}}\right)'+2 r {\sigma^{\mathrm{II}}} {\tau^{\mathrm{II}}} \Phi '\right)\non
&-\left({N^{\mathrm{II}}}\right)^2+2 r {\sigma^{\mathrm{II}}} {\tau^{\mathrm{II}}} \left({N^{\mathrm{II}}}\right)',\\
N^{\mathrm{II}}_{\mathrm{D}} =&-4
r {N^{\mathrm{II}}} \left(-r {\sigma^{\mathrm{II}}} {\tau^{\mathrm{II}}} \left({\eta^{\mathrm{II}}}\right)' \left(\Lambda '+\Phi '\right)+{\eta^{\mathrm{II}}} \left(-2 {\mu^{\mathrm{II}}}
\left(\Lambda '+\Phi '\right)-{\nu^{\mathrm{II}}} {\sigma^{\mathrm{II}}} \left(\Lambda '+\Phi '\right)\right.\right.\non
&\left.\left.+{\tau^{\mathrm{II}}} \left(r \left({\sigma^{\mathrm{II}}}\right)' \left(\Lambda
'+\Phi '\right)+{\sigma^{\mathrm{II}}}\left(4 r \Lambda '^2+4 \Phi '-2 r \Phi '^2+2 \Lambda ' \left(2+r \Phi '\right)+r \left(\Lambda ''+\Phi ''\right)\right)\right)\right)\right)\non
&+\left({N^{\mathrm{II}}}\right)^2 \left({\eta^{\mathrm{II}}}\right)^2+4 r^2 {\eta^{\mathrm{II}}} {\sigma^{\mathrm{II}}} {\tau^{\mathrm{II}}} \left({N^{\mathrm{II}}}\right)' \left(\Lambda '+\Phi '\right),\\
M^{\mathrm{II}}_{\mathrm{F}} =&-\frac{1}{2}{\kappa^{\mathrm{II}}}-\frac{1}{2} {\sigma^{\mathrm{II}}} \left(-2+e^{2 \Lambda } \lambda +2 r \Phi '\right) ,\\
N^{\mathrm{II}}_{\mathrm{F}} =&4 {N^{\mathrm{II}}} \left(-r
\sigma^{\mathrm{II}} {\tau^{\mathrm{II}}} \left(M^{\mathrm{II}}_\mathrm{F}\right)'+M^{\mathrm{II}}_\mathrm{F} \left(2 {\mu^{\mathrm{II}}}+{\nu^{\mathrm{II}}} {\sigma^{\mathrm{II}}}-{\tau^{\mathrm{II}}} \left(2 \sigma^{\mathrm{II}} \left(1+r \Lambda '\right)+r \left({\sigma^{\mathrm{II}}}\right)'\right)\right)\right) \non
&+\left({N^{\mathrm{II}}}\right)^2 {\kappa^{\mathrm{II}}}+4 r {M^{\mathrm{II}}_\mathrm{F}}{\sigma^{\mathrm{II}}} {\tau^{\mathrm{II}}}
\left({N^{\mathrm{II}}}\right)', \\
N^{\mathrm{II}}_{\mathrm{HK}} =&\left({N^{\mathrm{II}}}\right)^2 {\eta^{\mathrm{II}}}+4 r^2 {\sigma^{\mathrm{II}}} {\tau^{\mathrm{II}}} \left({N^{\mathrm{II}}}\right)' \left(\Lambda '+\Phi '\right)-4 r {N^{\mathrm{II}}}
\left(-2 {\mu^{\mathrm{II}}} \left(\Lambda '+\Phi '\right)-{\nu^{\mathrm{II}}} {\sigma^{\mathrm{II}}} \left(\Lambda '+\Phi '\right)\right.\non
&\left.+{\tau^{\mathrm{II}}} \left(r
\left({\sigma^{\mathrm{II}}}\right)' \left(\Lambda '+\Phi '\right)+{\sigma^{\mathrm{II}}} \left(2 r \Lambda '^2+3 \Phi '+\Lambda ' \left(3+2 r \Phi '\right)+r \left(\Lambda
''+\Phi ''\right)\right)\right)\right).
\end{align}

\subsection{Odd-parity for \texorpdfstring{$l\geq 2$}.}

The function $X$ in the \eq{oddmasterl2} takes the form
\begin{align}
X = e^{2\Lambda}\lambda-2r\Lambda'+2r\Phi'-2r^2\Lambda'\Phi'+2r^2{\Phi'}^2+2r^2\Phi'',
\end{align}
And the expression of $N_{\text{odd}}$ is given by
\begin{equation}
\label{Noddl2}
N_{\text{odd}} = -X+r\frac{ X'}{X} \left(-1+2 r \Lambda '+2 r \Phi '\right)+\left(2 r^2 \Lambda '^2+r \Phi '+2 r^2 \Phi '^2+r
\Lambda ' \left(1+4 r \Phi '\right)-2 \left(1+r^2 \Lambda ''+r^2 \Phi ''\right)\right).
\end{equation}

\subsection{The case for \texorpdfstring{$l=0$}.}

In this case, the functions $\mu_0$, $\nu_0$ and $\tau_0$ are given by
\begin{align}
\iota_0(r)&=-1+e^{2\Lambda}-r\Phi',\non
\sigma_0(r)&=2\left(1-e^{2\Lambda}\right)\Phi'+2r{\Phi'}^2+\Lambda'\left(-4+e^{2\Lambda}-4r\Phi'\right)+2r\Phi''.
\end{align}

\subsection{Odd parity for \texorpdfstring{$l=1$}.}
The function $X_1$ in the \eq{mastereqoddl1} takes the form
\begin{align}
X_1 = -2r\Lambda'+2r\Phi'-2r^2\Lambda'\Phi'+2r^2{\Phi'}^2+2r^2\Phi'',
\end{align}
And the expression of $N_{\text{odd}}^{(l=1)}$ is given by
\begin{equation}
N_{\text{odd}}^{(l=1)} = -X_1+r\frac{X_1'}{X_1} \left(-1+2 r \Lambda '+2 r \Phi '\right)+\left(2 r^2 \Lambda '^2+r \Phi '+2 r^2 \Phi '^2+r
\Lambda ' \left(1+4 r \Phi '\right)-2 \left(1+r^2 \Lambda ''+r^2 \Phi ''\right)\right).
\end{equation}
which is similar to the expression \eq{Noddl2}.

\section{The Degeneration Of \eqs{evenmasterEZ} and \meq{evenmasterRW}}
Considering that the metric of the background spacetime degenerates to the Schwarzschild spacetime, i.e.
\begin{equation}
ds^2=-\left(1-\frac{2M}{r}\right)dt^2+\left(1-\frac{2M}{r}\right)^{-1}dr^2+r^2(d\theta^2+\sin^2\theta d\varphi^2) ,
\end{equation}
where the Riemann curvature $R_{ab}$ and the scalar curvature $R$ all vanish. If we set
\begin{align}
\tilde{Z}^{(+)}=\frac{(r-2M)^2}{6M+r\tilde{\lambda}}Z^{(+)},
\end{align}
The  equations \eq{evenmasterEZ} and \meq{evenmasterRW} would become
\begin{align}
\label{degenemastereq}
\left[\frac{(r-2M)^2}{r^2}\frac{\pp^2}{\pp r^2}-\frac{\pp^2}{\pp t^2}+\left(\frac{r-2M}{r}\right)\frac{2M}{r}\frac{\pp }{\pp r}-V_{Z}\right]\tilde{Z}^{(+)}=S^{\text{w}}_Z,
\end{align}
where $ V_{Z} $ is the Zerilli effective potential
\begin{align}
V_{Z}=\left(\frac{r-2M}{r}\right)\left[\frac{72M^3+36M^2r\tilde{\lambda}+6Mr^2{\tilde{\lambda}}^2+r^3\tilde{\lambda}^2(\tilde{\lambda}+2)}
{r^3(6M+r \tilde{\lambda})^2}\right],
\end{align}
with the source term
\begin{align}
S^\text{w}_Z=\frac{r-2M}{2(6M+r\tilde{\lambda})}&\left[
\frac{96M^2r+r^3\tilde{\lambda}(\tilde{\lambda}-2)+2Mr^2(7\tilde{\lambda}-18)}{2(2M-r)(6M+r\tilde{\lambda})}E_\mathrm{A}+r^2\left(\frac{\pp}{\pp r}E_\mathrm{A}+\frac{\pp}{\pp t}E_\mathrm{D}\right)\right.\non
&\left.-\left(2+\tilde{\lambda}\right)\left(\frac{(6M+r\tilde{\lambda})}{2}E_\mathrm{D}-\left(r-2M\right)E_\mathrm{H} -\frac{(r-2M)}{2}E_\mathrm{K} \right)\right],
\end{align}
where $\tilde\lambda=\lambda-2=(l+2)(l-1)$. It is obvious that our result reduces to that of the Schwarzschild spacetime \cite{Thompson2016}.

\end{document}